\documentclass[11pt,a4paper]{article}
\usepackage{graphicx}
\usepackage{jheppub}
\pdfoutput=1

  \usepackage{amsmath}
  \usepackage{amsfonts}
  \usepackage{amssymb}

  \usepackage{graphicx}
  \usepackage{float}
  \DeclareGraphicsExtensions{.ps}

\title{Simulations of Cold Electroweak Baryogenesis: Quench from portal coupling to new singlet field}

	\author[b]{Zong-Gang Mou,}
	\author[a]{Paul M. Saffin,}
	\author[b]{Anders Tranberg}

	\affiliation[a]{School of Physics and Astronomy, University Park, University of Nottingham,\\ Nottingham NG7 2RD, United Kingdom}
	\affiliation[b]{Faculty of Science and Technology, University of Stavanger, 4036 Stavanger, Norway}

	\emailAdd{zonggang.mou@uis.no}
	\emailAdd{paul.saffin@nottingham.ac.uk}
	\emailAdd{anders.tranberg@uis.no}
	

\abstract{We compute the baryon asymmetry generated from Cold Electroweak Baryogenesis, when a dynamical Beyond-the-Standard-Model scalar singlet field triggers the spinodal transition. Using a simple potential for this additional field, we match the speed of the quench to earlier simulations with a ``by-hand" mass flip. We find that for the parameter subspace most similar to a by-hand transition, the final baryon asymmetry shows a similar dependence on quench time and is of the same magnitude. For more general parameter choices the Higgs-singlet dynamics can be very complicated, resulting in an enhancement of the final baryon asymmetry. Our results validate and generalise results of simulations in the literature and open up the Cold Electroweak Baryogenesis scenario to further model building.}

\begin{document}

\maketitle

\section{Introduction}
\label{sec:Intro}

The possibility of explaining the observed baryon asymmetry in the Universe as associated with the dynamics of electroweak symmetry breaking has a long history \cite{Kuzmin:1985mm,Cohen:1990it,Turok:1990in,Cohen:1993nk}. Underpinning this endeavour is the chiral anomaly in the electroweak sector of the Standard Model (SM), which establishes a relation between the Chern-Simons number of the SU(2) gauge fields and the baryon number of the fermions coupled to them \cite{tHooft:1976rip,tHooft:1976snw}. Any dynamical process whereby the Chern-Simons number changes in time will, therefore, be a candidate model for baryogenesis. 

Easily the most popular scenario on the table is to extend the SM by additional degrees of freedom\cite{Choi:1993cv,Huber:2006wf,Cheung:2012pg,Espinosa:2011ax,Damgaard:2015con,Alanne:2016wtx}, thereby allowing the symmetry breaking process to be a strongly first order finite temperature phase transition. To such a transition are associated bubbles of the low-temperature phase embedded in, and expanding into, the high-temperature background. These bubbles then grow, collide, and eventually the fields thermalise. As the broken-phase bubbles expand into the symmetric-phase, SM fermions scatter off the bubble wall leaving C and CP asymmetric densities in front of the progressing wall. These asymmetries bias the sphaleron transitions causing more baryons to be created than anti-baryons, and then the expanding bubble wall consumes this region of baryon over-density \cite{Cohen:1993nk,Morrissey:2012db}.

An alternative scenario that has received some attention is to instead postulate that interactions beyond the SM result in a cold state prior to symmetry breaking. Instead of a finite temperature phase transition driven by the expansion and cooling of the Universe, symmetry breaking could instead be a spinodal transition\footnote{In the context of the end of inflation, this is the process of tachyonic preheating.}\cite{Krauss:1999ng,GarciaBellido:1999sv,Rajantie:2000nj,Copeland:2001qw,Smit:2002yg}, triggered by the dynamics of Beyond-SM degrees of freedom. A number of realisations of this Cold Electroweak Baryogenesis scenario exist, and also a substantial body of work on computing the ensuing baryon asymmetry in different extensions of the SM \cite{Tranberg:2006ip,Konstandin:2011ds,Enqvist:2010fd,Mou:2015aia,Tranberg:2012qu,Tranberg:2012jp,Tranberg:2003gi,Tranberg:2006dg,vanderMeulen:2005sp,GarciaBellido:2003wd,Mou:2017atl,Mou:2017zwe}.

The most well-studied implementation involves the bosonic part of the electroweak sector, which comprises SU(2) and U(1) gauge fields as well as the Higgs field. In addition, CP-violation is introduced through a bosonic dimension six operator, which one would generically expect to arise from integrating out the fermionic degrees of freedom (see, however, \cite{Brauner:2011vb,Brauner:2012gu}). In a series of papers, the main features of this model were pinned down: that an asymmetry is created; that it is directly proportional to the dimensionless coefficient of the CP-violating term \cite{Tranberg:2006ip,Mou:2017zwe,Mou:2017atl}; and that the asymmetry is sensitively dependent on the Higgs mass (which has since been fixed by experiment) \cite{Tranberg:2006ip}.

The asymmetry generated is also very sensitive to the speed of the symmetry breaking quench. For very fast quenches, the asymmetry has the opposite sign compared to slow quenches \cite{Tranberg:2006dg}; the maximum asymmetry occurs for quenches lasting 10-20 $m_H^{-1}$ \cite{Mou:2017atl}. The asymmetry is also affected, by a factor of 2-3, by the inclusion of U(1) hypercharge fields in the dynamics in addition to the SU(2)-Higgs fields \cite{Mou:2017zwe}. 

In all previous simulations that included CP-violation explicitly, the symmetry breaking transition was triggered ``by hand" (see \cite{GarciaBellido:2003wd,GarciaBellido:2002aj,DiazGil:2008tf,DiazGil:2007dy} for dynamical symmetry breaking, but in a CP-even model). In these, the mass parameter $\mu$ in the Higgs potential was dialled to first provide a single minimum at $\phi=0$, and then the symmetry breaking was gradually switched on to give a potential minimum at the finite zero-temperature expectation value of 246 GeV. Ultimately, in a given model, the time-dependence of this mass parameter should be replaced by the dynamics of another degree of freedom, coupled to the Higgs field. Most likely the baryon asymmetry is model dependent, and the by-hand approach has the advantage of remaining agnostic about this. However, the dynamics of the new degree of freedom may introduce new effects and behaviours, badly captured by the non-dynamical triggering of the mass parameter, and that is what we explore in the following. 

In the present work, we will expand the model considered in \cite{Mou:2017atl} by adding a real scalar singlet with a simple quadratic potential. A quartic ``portal" coupling to the Higgs field provides dynamical symmetry breaking. We will see that, in a particular limit, we reproduce approximately the results of the by-hand approach, while for general choices of singlet parameters a number of other phenomena may arise. 

The paper is structured as follows: We start in section \ref{sec:quench} by introducing a simplified Higgs-singlet model, and discuss the types of behaviour one may expect from dynamical symmetry breaking. In section \ref{sec:model}, we then embed this two-scalar model into the electroweak sector of the Standard Model, giving a $SU(2)\times U(1)$-Higgs-singlet model with effective CP-violation. We review the observables and parameters in play, and describe the simulations to be performed. 
In section \ref{sec:quenchtimes} we present simulations of the case where the initial singlet energy is relatively small, and we match this limit to the by-hand method. In section \ref{sec:slow} we extend our simulations to also include higher energy singlet initial conditions, and describe the dynamics and asymmetry created in this case. As an aside, in section \ref{sec:ncs} we present and model the behaviour of the $N_{\rm cs,SU(2)}$ at intermediate and late times in the simulations. We conclude in section \ref{sec:conc}.

\section{Quench dynamics}
\label{sec:quench}

We will consider the bosonic part of the electroweak sector of the Standard Model, extended by a real scalar singlet. In later sections and in all of our simulations, we will include gauge fields and CP-violation, but setting aside these complications for the moment, we first consider the following action of two coupled scalar fields in order to better understand the dynamics of the process,
\begin{eqnarray}
S=- \int dt\, d^3x\Bigg[ \partial_\mu\phi^\dagger \partial^\mu\phi -\mu^2\phi^\dagger\phi + \lambda(\phi^\dagger\phi)^2+ \frac{1}{2}\partial_\mu\sigma\partial^\mu\sigma+\frac{m^2}{2}\sigma^2 + \xi^2 \sigma^2 \phi^\dagger\phi+V_0\Bigg],\nonumber\\
\end{eqnarray}
where $\sigma$ is a real gauge singlet and $\phi$ is the Higgs SU(2) doublet. The parameters $\lambda$ and $\mu$ are fixed by experiment to be $\mu=m_H/\sqrt{2}=88.4$ GeV and $\lambda=\mu^2/v^2=0.13$, where $v=246$ GeV is the Higgs vacuum expectation value (vev). The arbitrary constant $V_0=\mu^4/(4\lambda)$ is chosen so that the potential is zero in the global minimum. In addition, we have introduced two parameters, the BSM scalar's mass parameter $m$ and the scalar-Higgs coupling $\xi$. They are a priori free, although experimental collider constrains may be imposed, for instance on the singlet mass in the zero temperature vacuum \cite{Damgaard:2013kva}, 
\begin{eqnarray}
\label{eq:mmass}
m_\sigma^2 = m^2 + \xi^2 v^2.
\end{eqnarray}
Also, there are constraints on the mixing between the Higgs and the $\sigma$ (see for instance \cite{Damgaard:2015con}), but since in this model $\langle \sigma\rangle=0$, the mass matrix in the zero temperature vacuum is diagonal and there is no mixing. Mixing constraints would come into play, when allowing for a cubic coupling of the type $\sigma\phi^\dagger\phi$.

The structure of the potential is such that for $\sigma>\sigma_c=\mu/\xi$ the Higgs symmetry is unbroken ($\phi=0$), while for smaller $\sigma$ the Higgs field acquires a non-zero vev, tuned such that for vanishing $\sigma$ we reach the standard vacuum value for $\phi$, $\phi_{vac}=(0,v/\sqrt{2})$. The potential for the singlet has a single minimum at $\sigma=0$, and so the system will inevitably evolve to the usual Higgs vacuum, along with a vanishing vev for the singlet. We imagine that the conditions after inflation are such, that $\sigma(0)=\sigma_0> \sigma_c=\mu/\xi$, so the Higgs is initially in the symmetric phase, $\phi=0$. This may come about if the $\sigma$ is in fact the inflaton field itself, slow-rolling down some potential \cite{vanTent:2004rc}. Or, if it is a spectator field, one may argue that stochastically it will have a non-zero value at the end of inflation \cite{Starobinsky:1986fx,Nakao:1988yi,Stewart:1991dy,Starobinsky:1994bd,Enqvist:2012xn}.

We expect that the singlet $\sigma$ is homogeneous as a result of the inflationary expansion. This means that the initial condition can be described by $\sigma_0$ and $\dot{\sigma}(0)=\dot{\sigma}_0$. Without loss of generality, we may set $\dot{\sigma}_0=0$, since any non-zero value at some $\sigma_0$ corresponds to zero initial speed but from some other (larger) $\sigma_0$. Since $\sigma$ is initialised at a finite value, as $\sigma$ rolls down towards zero, symmetry breaking and the spinodal transition is triggered at the critical value $\sigma_c$.
For this analysis we will ignore the expansion of the Universe, since for electroweak energies the Hubble time $H^{-1}$ is much longer than the time scale of the dynamics $m_W^{-1}$.

In our model, we are left with three free parameters: $m$, $\xi$ and $\sigma_0$, and in principle one could simply compute the baryon asymmetry, scanning through these. However, for reasons to become clear below, we will reparametrise this 3-dimensional space. We first express $\sigma_0$ in terms of $\sigma_c$ as  $\sigma_0=A\mu/\xi$, which defines the dimensionless parameter $A$. Second, we introduce the total initial energy and use it to define $n$
\begin{eqnarray}
\label{eq:n_energy}
E_{\rm tot} = V_0 + \frac{m^2}{2}\sigma_0^2 = V_0\left(1+\frac{m^2}{m_H^2}\frac{4A^2\lambda}{\xi^2}\right)\equiv V_0\left(1+\frac{1}{n^2}\right), \qquad n= \sqrt{\frac{\xi^2}{4A^2\lambda}}\frac{m_H}{m}.\nonumber\\
\end{eqnarray}
This allows us to scan the parameter space in terms of the physically more intuitive dimensionless parameters $m_H/m$, $n$ and $A$. First, we will explain how these quantities are constrained by the scenario, and how they are related to the by-hand quench of \cite{Mou:2017atl}.

\subsection{Simple constraints}
\label{sec:constraint}

\begin{enumerate}
\item We will be initialising the Higgs field with free-field quantum vacuum fluctuations, to seed the spinodal growth (see \cite{Rajantie:2000nj,GarciaBellido:2002aj,Smit:2002yg}). These depend on the initial mass of the Higgs field which is then
\begin{eqnarray}
\mu_{\rm eff}^2(0) = \xi^2 \sigma_0^2-\mu^2 = (A^2-1)\mu^2 = \frac{1}{2}\left(A^2-1\right)m_H^2. 
\end{eqnarray}
In \cite{Mou:2017atl}, we used $A^2=2$, corresponding to $\mu_{\rm eff}^2(0)=\mu^2$. We will do the same below, although in principle one may choose any value $A>1$. 

\item Secondly, a basic requirement for Cold Electroweak Baryogenesis is that the temperature after the transition and thermalisation should be less than the equilibrium electroweak phase transition temperature of $\simeq 160$ GeV \cite{Kajantie:1996mn,Karsch:1996yh,Aoki:1996cu,Gurtler:1997hr,Laine:1998jb,DOnofrio:2014rug,Laine:2015kra}. Assuming that the singlet $\sigma$ counts as a relativistic degree of freedom after the transition, this means that distributing all the available energy, we have
\begin{eqnarray}
V_0\left(1+\frac{1}{n^2}\right)=\frac{\pi^2}{30}g^* T^4,
\end{eqnarray}
with an effective number of degrees of freedom $g^*=16+2+1+\frac{7}{8}(18+60)=87.25$, as the top quark and massive vector bosons are heavier than the assumed temperature scale. Requiring that $T<160$ GeV, using $m_H=125$ GeV and $\lambda=0.13$, we find $n>0.08$, or equivalently $E_{\rm tot}<158 V_0$. In the limit $n\rightarrow \infty$, $T=45$ GeV. We note that in the simulations, only 13 degrees of freedom are present, so that the final temperature is somewhat higher. But the time-scales of the simulations will not allow us to reach thermal equilibrium. 
\item Thirdly, we can make the connection to the by-hand transition of \cite{Mou:2017atl}, where instead of a dynamically evolving field $\sigma$, the Higgs field experienced a mass quench through the replacement
\begin{eqnarray}
-\mu^2\rightarrow\mu^2_{\rm eff}(t)=\mu^2\left(1-\frac{2t}{\tau_q}\right),\quad 0<t<\tau_q,
\end{eqnarray}
and $-\mu^2$ for $t>\tau_q$. The quench is then parametrized by a quench time $\tau_q$. We note that $\mu_{\rm eff}^2(0) = +\mu^2$, corresponding to the choice $A^2=2$ made above. We may define a quench speed as the dimensionless speed at the time where $\mu_{\rm eff}^2$ goes through zero and symmetry breaking is triggered:
\begin{eqnarray}
u=\frac{1}{2\mu^3}\frac{d \mu^2_{\rm eff}(t)}{dt}|_{\mu_{\rm eff}^2=0} = -\frac{1}{\mu\tau_q}.
\end{eqnarray}
Similary, we may compute this for the dynamical case with $\mu_{\rm eff}^2(t) = \xi^2\sigma^2(t)-\mu^2$
\begin{eqnarray}
u=\frac{1}{2\mu^3}\frac{d\mu^2_{\rm eff}(t)}{dt}|_{\mu_{\rm eff}^2=0} = \frac{1}{\mu}\frac{\dot{\sigma}_c}{\sigma_c},
\end{eqnarray}
with $\sigma_c=\mu/\xi$. In the limit where only the quadratic $\sigma$-potential contributes, 
\begin{eqnarray}
\sigma(t) = \frac{A\mu}{\xi}\cos(mt)\rightarrow u = -\frac{m}{\mu}\sqrt{A^2-1}.
\end{eqnarray}
Hence, for $A^2=2$, it is tempting to make the identification $\tau_q=m^{-1}$. Once the Higgs field starts to evolve away from zero, the true potential of $\sigma$ is somewhat different, and so this identification is not exact. As will see below, there is a proportionality constant of order one.
\end{enumerate}

Since we are mostly interested in the quench time dependence, we will in the following set $A^2=2$, and vary $m_H/m$ for a few values of $n$. For example, in section \ref{sec:quenchtimes} we will examine $n=8$, corresponding to a very ``cold" $\sigma$, where the energy in the system is simply $1.02 V_0$, finding that in this case the behaviour and baryon asymmetry produced is very similar to the by-hand quench. In section \ref{sec:slow} we consider fast quenches, $m_H/m=4$, for different values of $n$ in the interval $1\to8$.

\section{The quenched $SU(2)\times U(1)$-Higgs-singlet model with CP-violation}
\label{sec:model}

After having surveyed the quench mechanism, we can now embed the two-field model in the full electroweak sector of the Standard Model. This is composed of a Higgs doublet coupled to SU(2) and U(1) gauge fields, and in addition the new scalar singlet. Instead of adding the entire fermion sector dynamically \cite{Saffin:2011kn}, we will imagine having integrated out all the other degrees of freedom, and that any SM and BSM CP-violation is retained in an effective dimension-six term \cite{GarciaRecio:2009zp,Hernandez:2008db,Brauner:2011vb,Brauner:2012gu}. The classical action reads
\begin{eqnarray}
\label{eq:S_EW}
S=- \int dt\, d^3x\Bigg[ \frac{1}{2}\textrm{Tr}\,W^{\mu\nu}W_{\mu\nu} +\frac{1}{4} B^{\mu\nu}B_{\mu\nu}+\frac{3\delta_{\rm cp}g^2}{16\pi^2 m_W^2}\phi^\dagger\phi \textrm{Tr}\,W^{\mu\nu}\tilde{W}_{\mu\nu} \nonumber\\\qquad
+ (D_\mu\phi)^\dagger D^\mu\phi -\mu^2\phi^\dagger\phi + \lambda(\phi^\dagger\phi)^2+ \frac{1}{2}\partial_\mu\sigma\partial^\mu\sigma+\frac{m^2}{2}\sigma^2 + \frac{1}{2}\xi \sigma^2 \phi^\dagger\phi\Bigg].\nonumber\\
\end{eqnarray}
The field strength tensors are $W_{\mu\nu}$ for SU(2) and $B_{\mu\nu}$ for U(1).  The gauge couplings are $g$ and $g'$, respectively, and we have the Higgs self-interaction $\lambda$ and mass parameter $\mu$ as before. The latter two can be replaced by the observed values of the Higgs vev and Higgs mass
\begin{eqnarray}
m_H^2 = 2\mu^2 = 2\lambda v^2.
\end{eqnarray}
The covariant derivative $D_\mu$ is given by
\begin{eqnarray}
D_\mu\phi = \left(\partial_\mu +i \frac{1}{2}g'B_\mu-i gW_\mu^a\frac{\sigma_a}{2}\right)\phi,
\end{eqnarray}
with the U(1) gauge field $B_\mu$ and the SU(2) gauge field denoted by $W_\mu$. We have used that the Higgs field hypercharge $Y=-1/2$.

This leaves, as before, two parameters in the Higgs-scalar sector, $m$, $\xi$ as well as the $\sigma$ initial condition $\sigma_0$. We also have the parameter determining the strength of the CP-violation, $\delta_{\rm cp}$. The dependence of the baryon asymmetry on $\delta_{\rm cp}$ has been determined in a series of works \cite{Mou:2017zwe,Mou:2017atl}, with the result that it is linear for reasonably small values $\delta_{\rm cp}\lesssim 10$, as we will confirm below. For numerical reasons (to see the numerical signal clearly), it is convenient to use a fairly large value of $\delta_{\rm cp}$, and we use $3\delta_{\rm cp}=20$ unless explicitly stated otherwise. We also use the physical values $m_H=125$ GeV, $v=246$ GeV, $m_W=80$ GeV and $m_Z=91$ GeV, therefore $g=0.65$ and $g'=0.35$.

\subsection{Simulations of Cold Electroweak Baryogenesis}
\label{sec:coldbaryogenesis}

Details of Cold Electroweak Baryogenesis may be found elsewhere \cite{Smit:2002yg}, but, in short, the mechanism is based on the fact that as a Higgs symmetry-breaking is triggered, Higgs field modes with $k<\mu$ become unstable and grow exponentially, a process known as tachyonic preheating or spinodal decomposition. This is a strongly out-of-equilibrium process, with all the power in the infra-red (IR), and in the presence of CP-violation a net baryon asymmetry is created. 

In our strictly bosonic model, we invoke the chiral anomaly to make the identification
\begin{eqnarray}
B(t)-B(0) = 3\left[N_{\rm cs,SU(2)}(t)-N_{\rm cs,SU(2)}(0)\right],
\end{eqnarray}
where $N_{\rm cs,SU(2)}$ is the SU(2) Chern-Simons number\footnote{There is also a contribution from the $U(1)$ Chern-Simons number, but it does not lead to a permanent change in baryon number, as it is zero in the vacuum/at late times.} \cite{tHooft:1976rip,tHooft:1976snw}. In the specific context of Cold Electroweak Baryogenesis, the anomaly was explicitly confirmed in simulations with dynamical fermions \cite{Saffin:2011kn}. In addition, it turns out that because of the violent nature of the transition, and the rather long thermalisation times, it is convenient to make the further identification
\begin{eqnarray}
N_{\rm cs,SU(2)}(t)-N_{\rm cs,SU(2)}(0)\simeq N_{\rm w}(t)-N_{\rm w}(0),
\end{eqnarray}
where $N_{\rm w}$ is the Higgs field winding number. The reason is that $N_{\rm w}$ is an integer (up to lattice discretization errors), and therefore a much cleaner observable than $N_{\rm cs,SU(2)}$. Also, whereas $N_{\rm cs,SU(2)}$ oscillates for a long time, $N_{\rm w}$ settles very early in the simulation. At very late times (as we checked) $N_{\rm cs,SU(2)} \rightarrow N_{\rm w}$. We will discuss the behaviour of $N_{\rm cs,SU(2)}$ in some detail in section \ref{sec:ncs}. Hence, although in our simulations we monitor several observables, including $N_{\rm cs,SU(2)}$, we will ultimately infer $B = 3N_{\rm w}$. 

On a more technical note, we will follow the procedure in \cite{Tranberg:2010af,Mou:2015aia}, and average our observables over an explicitly CP-even ensemble of random classical initial conditions. This is achieved by taking pairs of initial conditions, so that for every realisation we also include its CP-conjugate in the ensemble. This implies that for $\delta_{\rm cp}=0$, the baryon asymmetry is identically zero. In this work, the ensembles count 200-400 such CP-conjugate pairs.

From a simulation perspective, we need to have a lattice resolution fine enough to convincingly represent the UV dynamics and compute observables accurately (notably the Higgs winding number). We use a lattice spacing $a$, so that $am_H=0.375$. We also need a large enough spatial volume such that the relevant dynamics fits inside the box. This requires 
that the linear size of the lattice, $L$, is big enough, and we use $Lm_H=24$. This also ensures that the number of unstable tachyonic modes is large enough to mimic a continuum of modes. Finally, we must ensure that also the dynamics of the $\sigma$ field is well contained. Trivially, $Lm= 24 (m/m_H)$, and even for  $m/m_H\simeq 4$ one may worry that this is too small.  Fortunately, the mass of the $\sigma$ field is not $m$ once the tachyonic transition is triggered but rather given by eq. (\ref{eq:mmass}), allowing us to rewrite
\begin{eqnarray}
\label{eq:volume}
Lm_\sigma = Lm_H\frac{m}{m_H}\sqrt{1+4n^2}.
\end{eqnarray}
Hence for $n=8$, even $m_H/m$ up to 30-40 is probably reliable. For $n=1$, we should not trust $m_H/m$ larger than around 6. We have tested somewhat larger volumes to confirm these estimates give the correct scales at which our dynamics converges. We also see that the masses, in lattice spacing units, follows a similar relation
\begin{eqnarray}
\label{eq:spacing}
am_\sigma = am_H\frac{m}{m_H}\sqrt{1+4n^2}.
\end{eqnarray}
With $am_H=0.375$ and $n=8$, we find $am_\sigma\simeq 6\, m/m_H$, at least at the end of the simulation when the $\sigma$ field settles. Our fastest quench of $m_H/m=4$ therefore comes with some reservations, although we will see that the results are consistent with other $m_H/m$. Conversely, for $n=1$ and $m_H/m$, $am_\sigma<am_H$, and all is well under control.

\section{Cold quenches, $n=8$}
\label{sec:quenchtimes}

\begin{figure}
\begin{tabular}{lr}
\hspace{-1.5cm}
\includegraphics[width=0.6\textwidth]{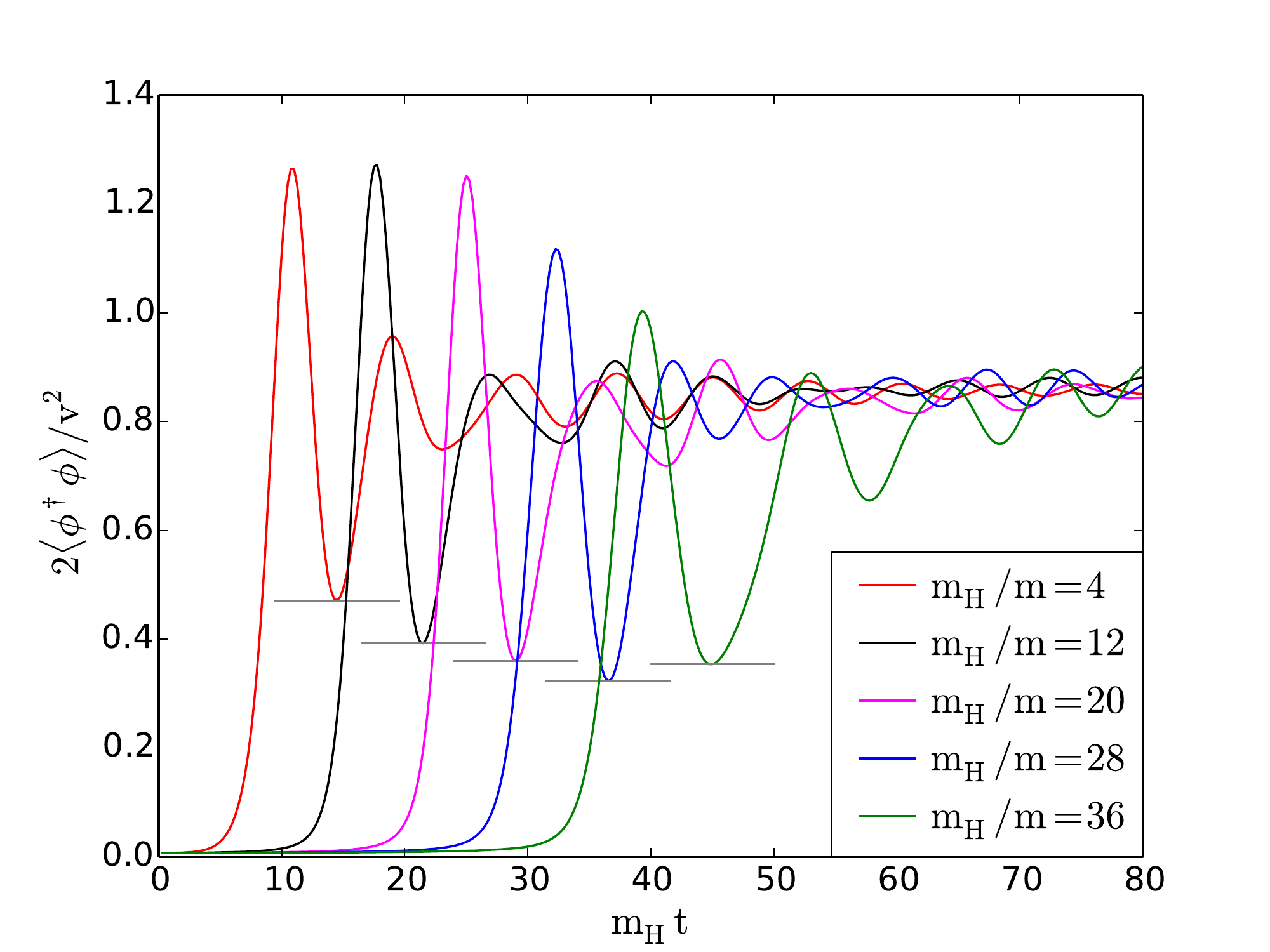} & \hspace{-1.2cm}
\includegraphics[width=0.6\textwidth]{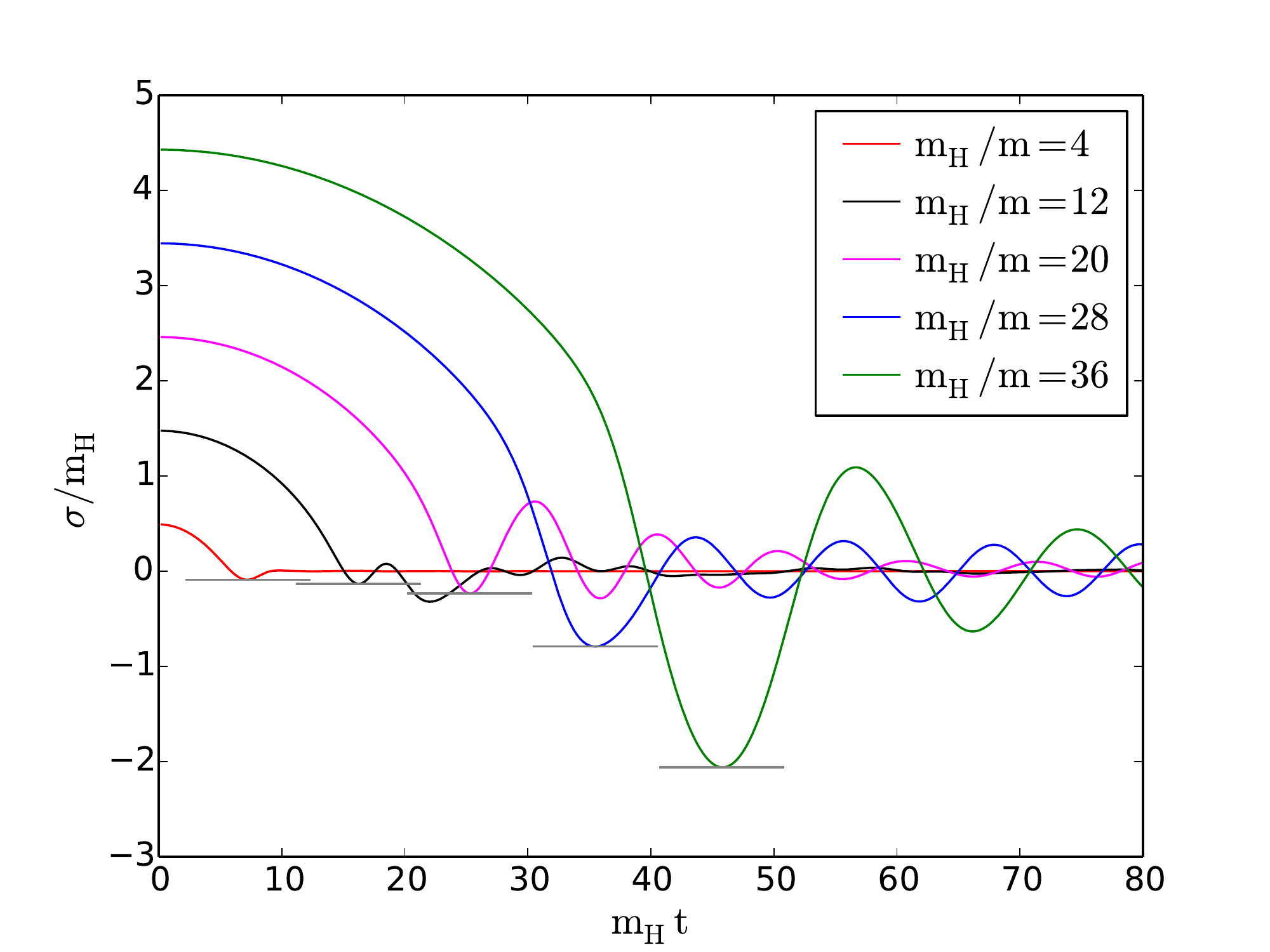}
\end{tabular}
\caption{Left: The average Higgs field in time, for $n=8$ and different quench rates $m_H/m$. Black horizontal lines indicate the first Higgs minimum, used to define the quench time $T_1$. Right: The $\sigma$ field for the same simulations.}
\label{fig:match1}
\end{figure}

\begin{figure}
\begin{center}
\includegraphics[width=0.6\textwidth]{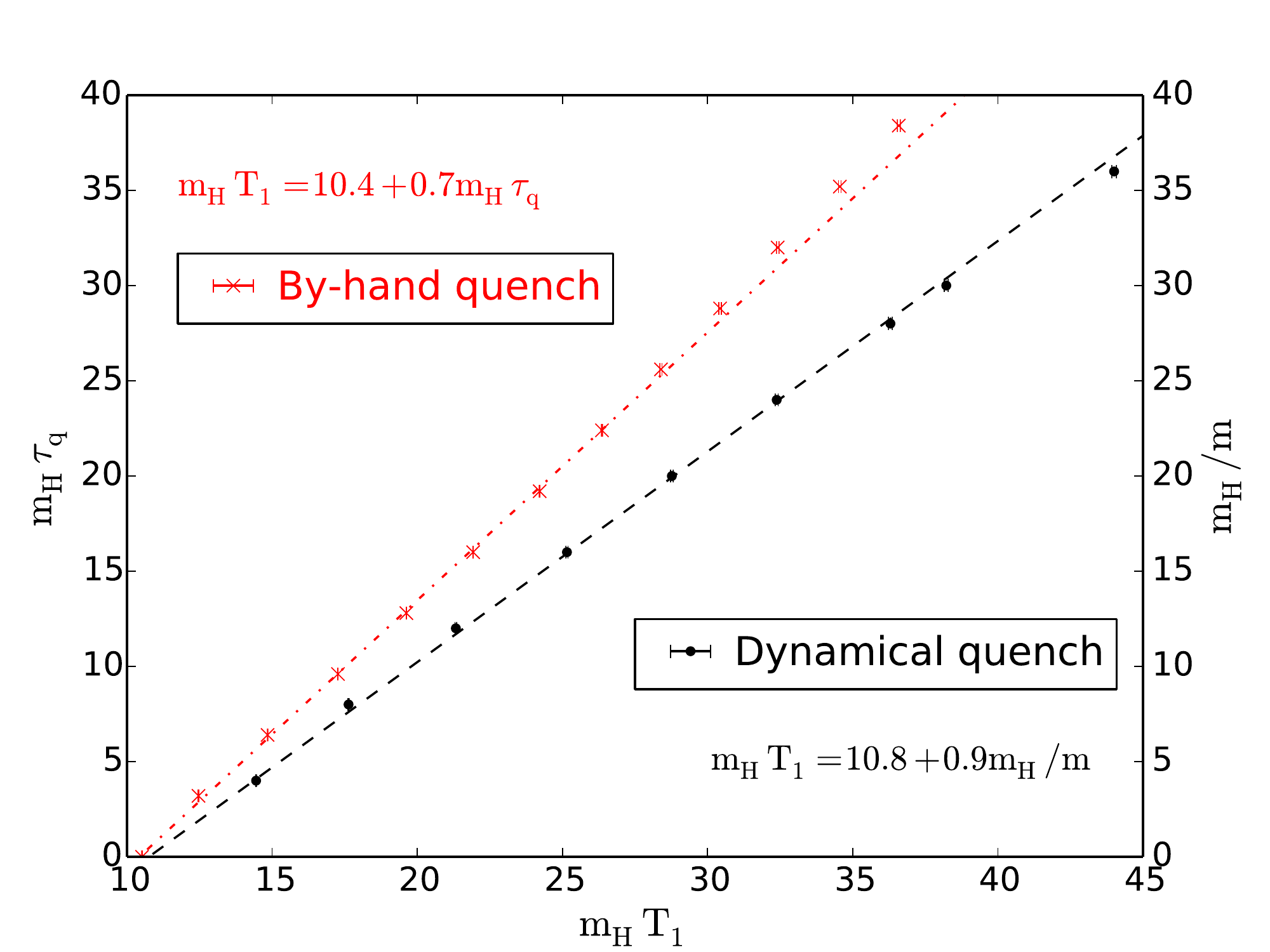}
\end{center}
\caption{The relation between quench time $T_1$ and $\tau_q$ and $m^{-1}$, respectively. Even when the mass flip is instantaneous, the Higgs takes a finite time (about 10 $m_H^{-1}$) to complete the transition. }
\label{fig:match2}
\end{figure}

We first consider the case where there is little energy in the $\sigma$-potential, and take $n=8$ to represent the large-$n$ limit, giving a total energy of $1.02\times V_0$. We now introduce a definition of the ``quench time" $T_1$, as the time it takes for the Higgs field to reach its first minimum in its oscillations as shown in Fig.\ref{fig:match1} (left panel). Also in Fig.\ref{fig:match1} (right panel), we show the $\sigma$ field in the same simulations. As discussed above, in previous work \cite{Mou:2017atl}  the transition was triggered by flipping the sign of the Higgs mass coefficient over a timescale $\tau_q$. We may use the same definition for the duration $T_1$ in that case. In Fig. \ref{fig:match2}, we show the $m_HT_1$ as a function of $m_H/m$ (right vertical axis) and as a function of $m_H\tau_q$ (left vertical axis). We see that there is clear proportionality, and that the relation may be written
\begin{eqnarray}
\tau_q \simeq 1.3\,m^{-1}.
\end{eqnarray}

\begin{figure}
\begin{tabular}{lr}
\hspace{-1.5cm}
\includegraphics[width=0.6\textwidth]{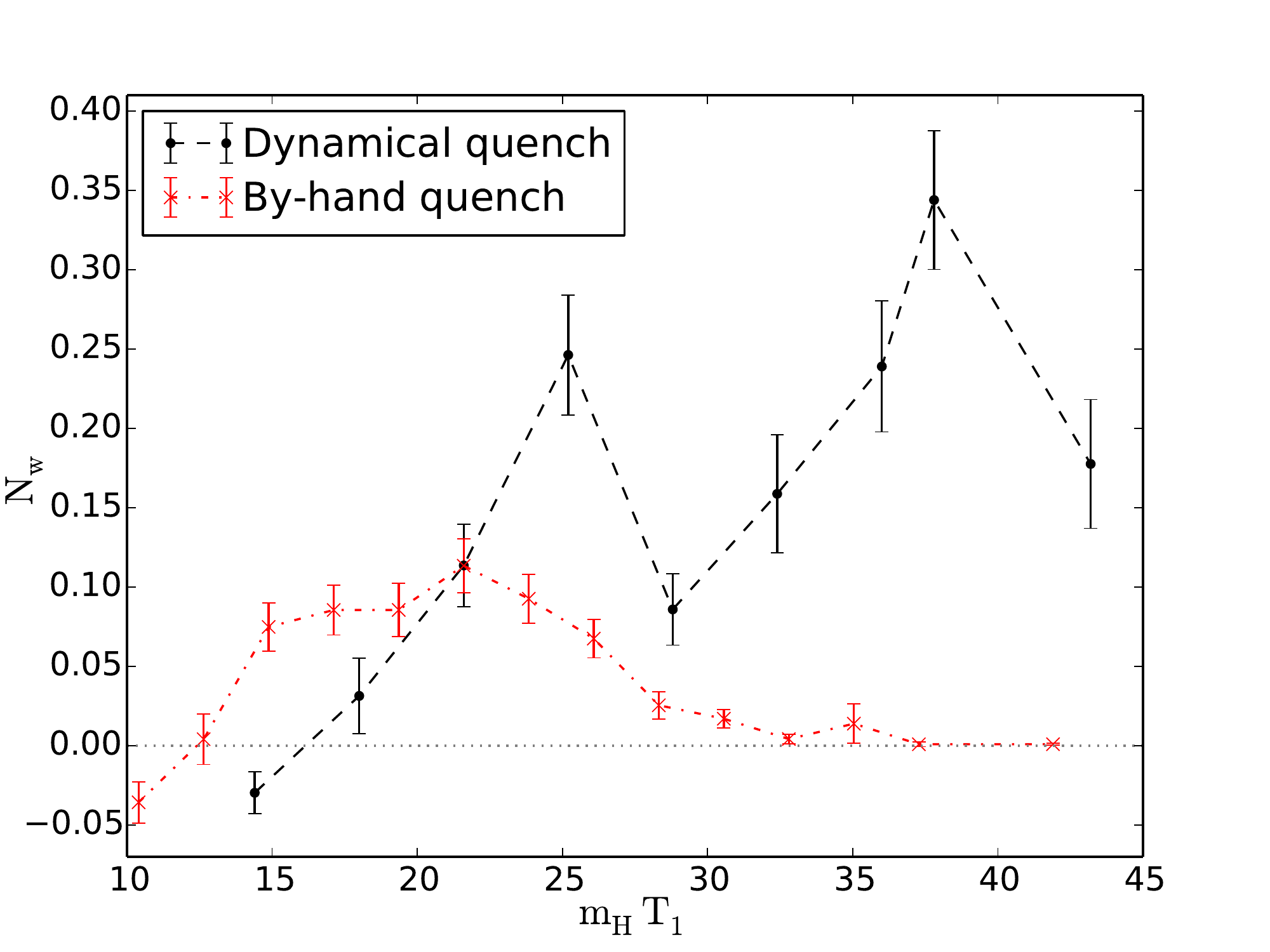} & \hspace{-1.2cm}
\includegraphics[width=0.6\textwidth]{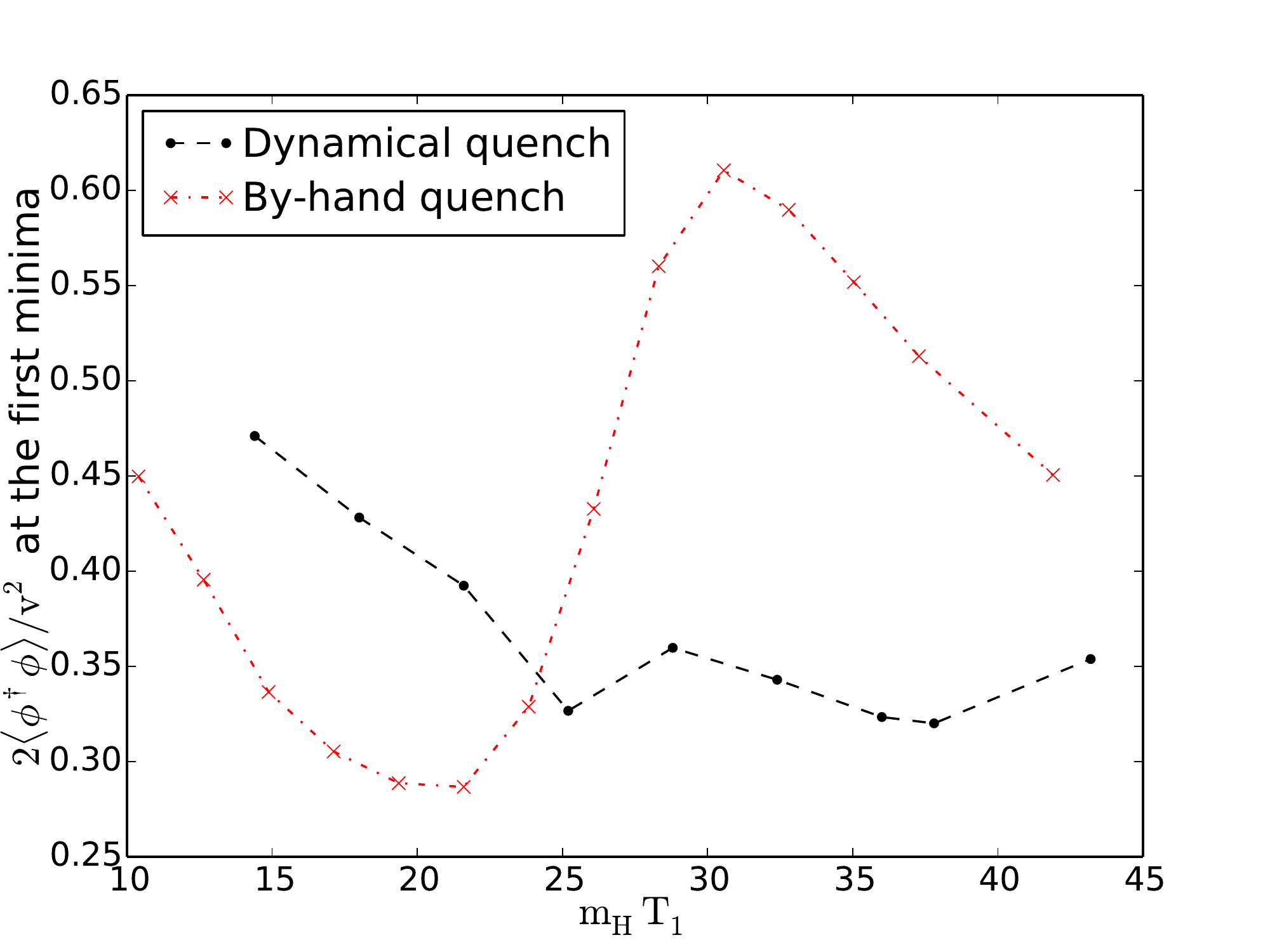}
\end{tabular}
\caption{Left: The final asymmetry (in $N_{\rm w}$) for the dynamical (black) and by-hand (red) simulations. Right: The value of the Higgs field (squared) at the first minimum. Note the strong correlation between a low Higgs minimum and a large asymmetry.}
\label{fig:asym_match}
\end{figure}

Having calibrated the dynamical-$\sigma$ simulations against the by-hand simulations we can proceed with computing our primary observable $\langle N_{\rm w}\rangle$ as a proxy for the baryon asymmetry, and uncover the consequences of allowing the electroweak symmetry to break dynamically, rather than quenching by hand. In Fig. \ref{fig:asym_match}  (left) we show the asymmetry in $\langle N_{\rm w}\rangle$ for $n=8$ dynamical quench simulations, as well as for by-hand simulations, where we have rescaled to $m_HT_1$ to make the comparison. 

We see that there is a qualitative agreement, in the sense that for very fast quenches, the asymmetry is negative and of order $\langle N_{\rm w}\rangle=0.03$; while for slower quenches the asymmetry becomes positive with one (by-hand) or two (dynamical) maxima. The maximum by-hand asymmetry is around $\langle N_{\rm w}\rangle=0.1$. For the dynamical simulations, the asymmetry peaks at values of $\langle N_{\rm w}\rangle=0.25$ and $0.35$.
This suggests that the by-hand simulations, in particular for fast quenches, are really the large-$n$ limit of dynamical quench simulations. The limit where the total energy is essentially the initial Higgs potential. 

The peak structure was observed before for the by-hand quench \cite{Tranberg:2006dg,Mou:2017atl} and can be traced to the larger abundance of local zeros of the Higgs field, allowing Higgs winding to occur. This, in the presence of CP-violation, leads to a baryon asymmetry. In Fig. \ref{fig:asym_match} (right) we clearly see a strong correlation between the obtained asymmetry and the value of the average Higgs field at the first minimum (where we also define $T_1$). A low minimum corresponds to many local Higgs zeros.

An explanation why there are more Higgs zeros at certain values of the quench time is more subtle. Qualitatively, it follows from the shape of the Higgs potential at the time of the first Higgs minimum, and the speed of the quench. In essence, it is a question of whether the Higgs field can ''slosh back up" the Higgs potential, either because it has large speed (by-hand peak and first dynamical peak), or because the potential is shallower at that moment (second dynamical peak). 

\begin{figure}
\begin{tabular}{lr}
\hspace{-1.5cm}
\includegraphics[width=0.6\textwidth]{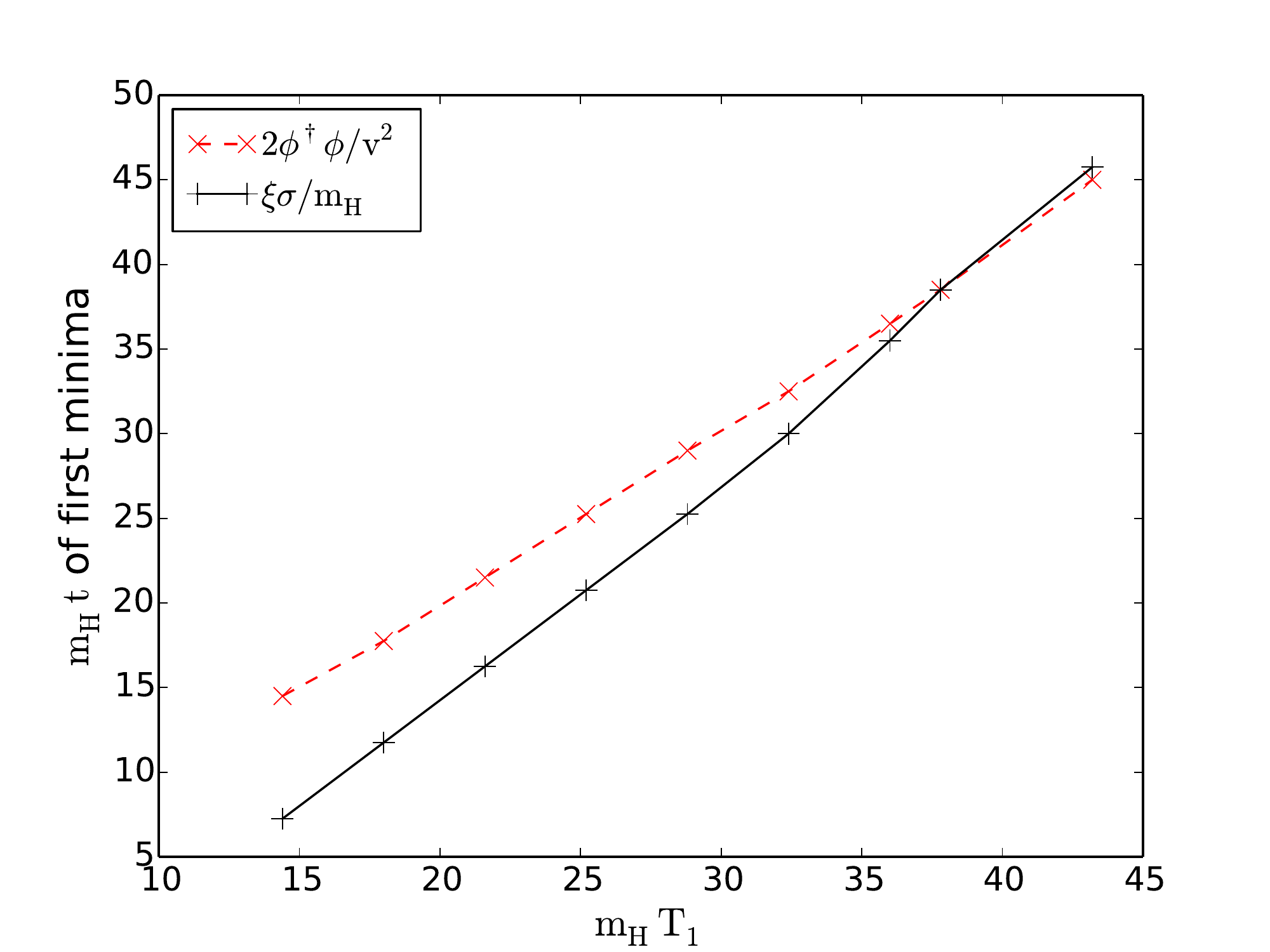} & \hspace{-1.2cm}
\includegraphics[width=0.6\textwidth]{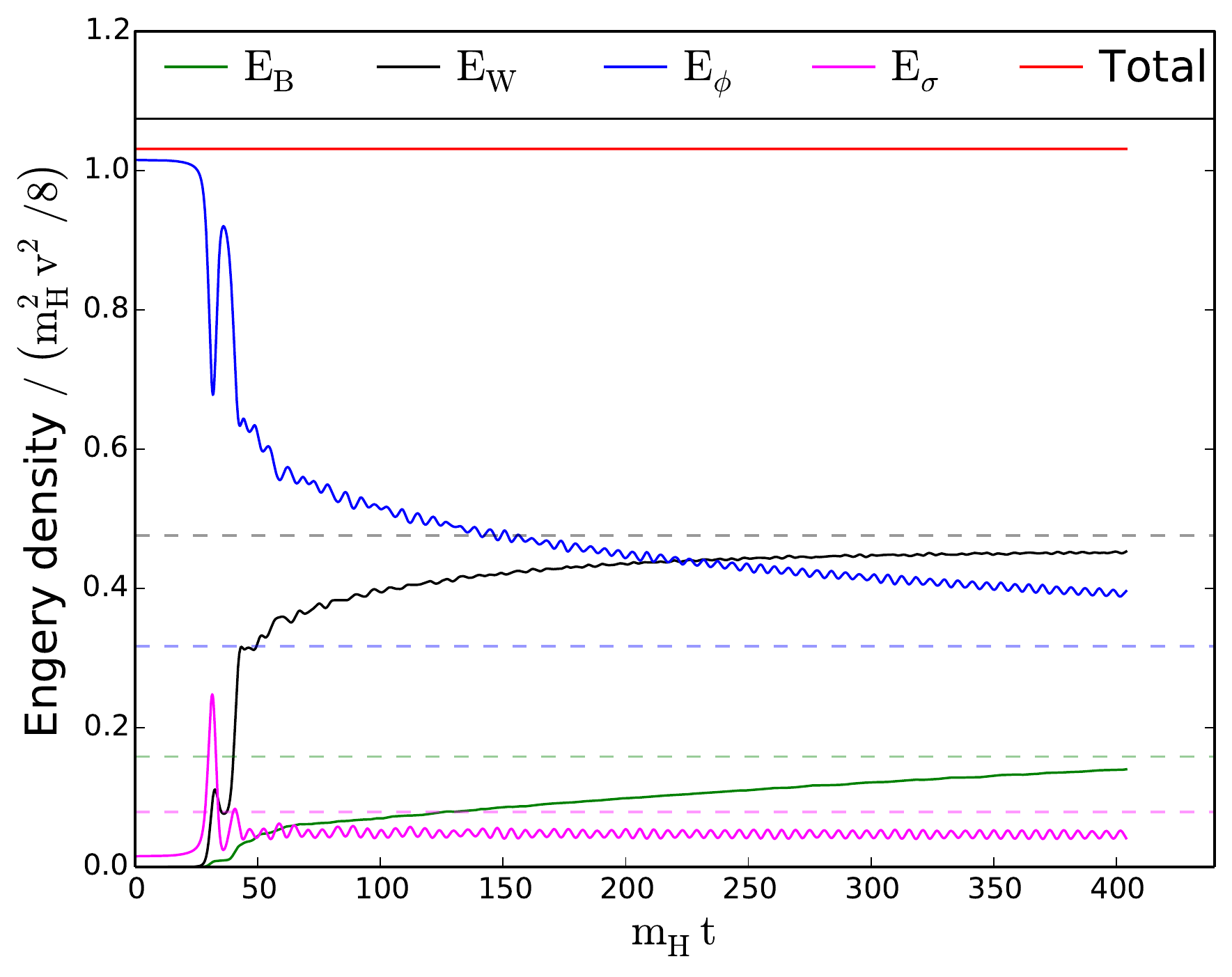}
\end{tabular}\caption{Left: The time of the first Higgs minimum and first $\sigma$ minimum for different quench times. Right: The energy components in an $n=8$, $m_H/m=28$ simulation ($B$ is $U(1)$ gauge field, $W$ is $SU(2)$. $\phi$ and $\sigma$ the two scalar fields. Dashed lines denote the expected equipartition asymptotics.}
\label{fig:energy}
\end{figure}

As concerns the latter, Fig. \ref{fig:energy} (right) shows the time of the first Higgs maximum and the first $|\sigma|$ maximum as a function of $m_H/m$. The second dynamical peak in the asymmetry occurs precisely when the two coincide ($m_H/m=30$) which turns out also to be when the maximum $|\sigma|$ is largest. This corresponds to the Higgs potential being shallower than in the global minimum, and this generates many Higgs zeros and hence the second dynamical peak. Had $|\sigma|$ been even larger $>\sigma_c$, the symmetry of the potential would have been restored, and the transition halted. 

Accepting the matching in terms of quench time $T_1$, one may conclude that the inclusion of dynamical symmetry breaking makes the maximum asymmetry occur at somewhat slower quenches. But that the negative sign of the asymmetry at the fastest quenches is a robust prediction, and not an artefact of triggering the quench by-hand.

\subsection{Where does the energy go?}

Another point to make is that in the by-hand simulations, energy is extracted from the system, because of the time-dependence of $\mu^2$. It is easy to see that the energy loss is 
\begin{eqnarray}
\Delta E = -\frac{\mu^2}{\tau_q}\int_0^{\tau_q} dt\, d^3x\, \phi^\dagger\phi(x,t),
\end{eqnarray}
which for the quenches in \cite{Mou:2017atl} was as much as 60\%.
As a result of a different effect, energy is also extracted from the gauge-Higgs system in a large-$n$ dynamical quench. At late times, energy equipartition assigns a certain fraction of the total energy to the $\sigma$ degree of freedom. Simple counting of all the degrees of freedom reveals that $1/13$ ends up in the $\sigma$ field. In Fig. \ref{fig:energy} we show the time evolution of the different energy components, with dashed line indicating their expected asymptotic values. Note that the distribution between gauge and Higgs degrees of freedom may have some gauge dependence. In this incomplete, temporal gauge choice, it seems that the energy from the shared modes is mostly in the Higgs field (4 d.o.f. rather than just 1 Higgs mode) and not in the gauge field (massless fields, 6 d.o.f., rather than massive, 9 d.o.f.). We expect 1/13 of the energy to go into the  $\sigma$ field. Because $n=8$, the initial energy in the $\sigma$ field is less than its equipartition value, and so qualitatively (this effect is not quench-time dependent), for this $n$, the effect of including the dynamical quench is not to add, but to extract energy from the gauge-Higgs system. This adds to the understanding why the by-hand approach works reasonably well. 

\section{Warmer, and fast quenches: $n=1\to8$, $m_H/m=4\;(m_H\tau_q\simeq 5)$}
\label{sec:slow}

We now proceed to consider other values of $n$, for which the results depart significantly from the by-hand simulations. Smaller $n$ means that more energy is present in the system, as we see from (\ref{eq:n_energy}), and initially it is stored in the initial potential energy of the $\sigma$ field. Hence, as $n$ reduces we expect the dynamics to inject more and more energy into the SM sector. Related to our prior discussion of equipartition, the $\sigma$ has more initial energy than its fair share of $1/13$, when $n<\sqrt{12}$. But we have also seen that at intermediate times, the energy distribution may deviate substantially from equipartition. 

We will restrict ourselves to the range $n=1\to8$, corresponding to energies between $V_0$ and $2V_0$. Considering again all the degrees of freedom of the whole SM, this in turn corresponds to reheating temperatures of $T_{\rm reh}=54\to45$ GeV, so is still deep in the broken phase. 

\begin{figure}
\begin{tabular}{lr}
\hspace{-1.5cm} \vspace{-0.15cm}
\includegraphics[width=0.6\textwidth]{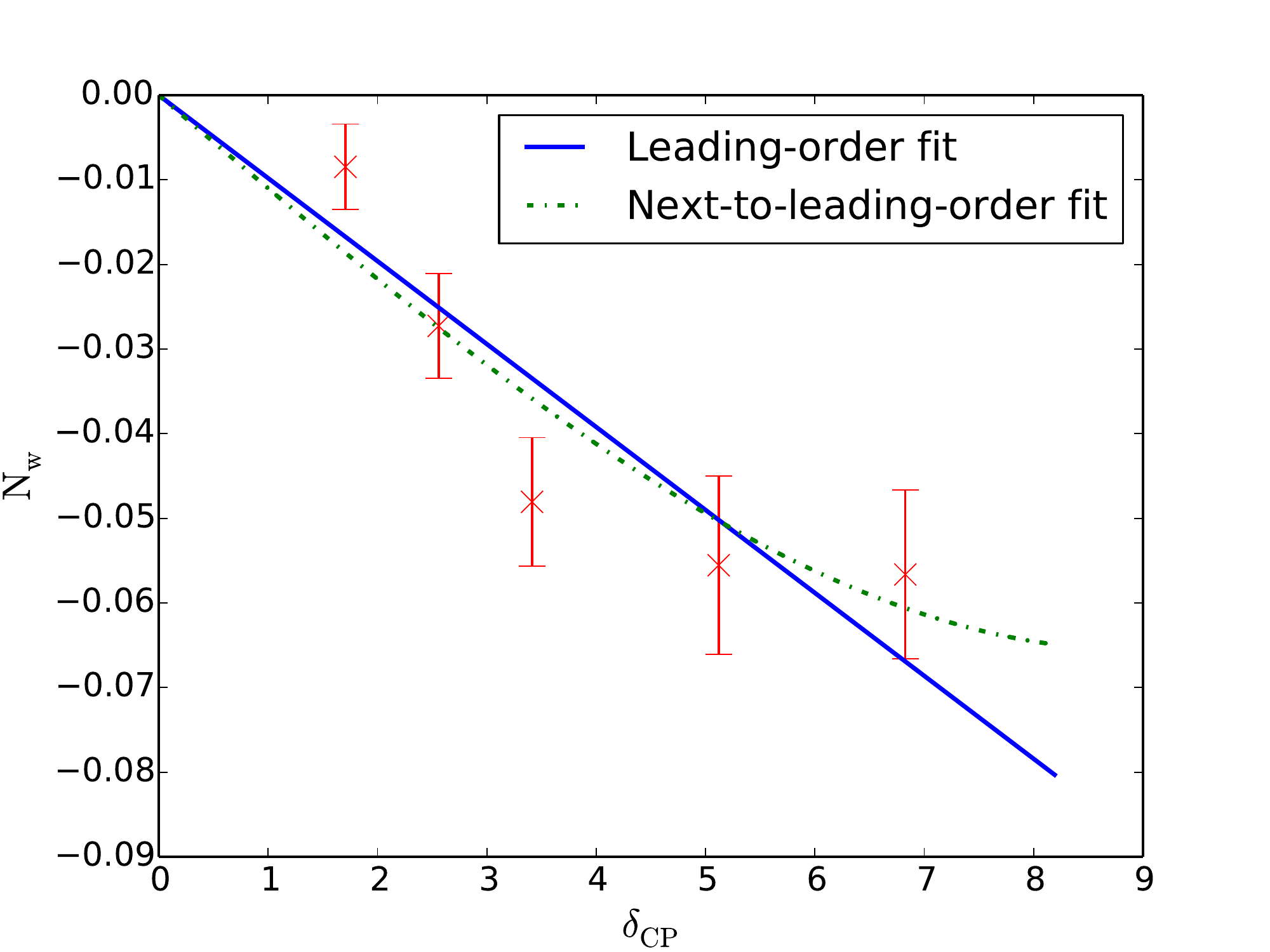} & \hspace{-1.1cm}
\includegraphics[width=0.6\textwidth]{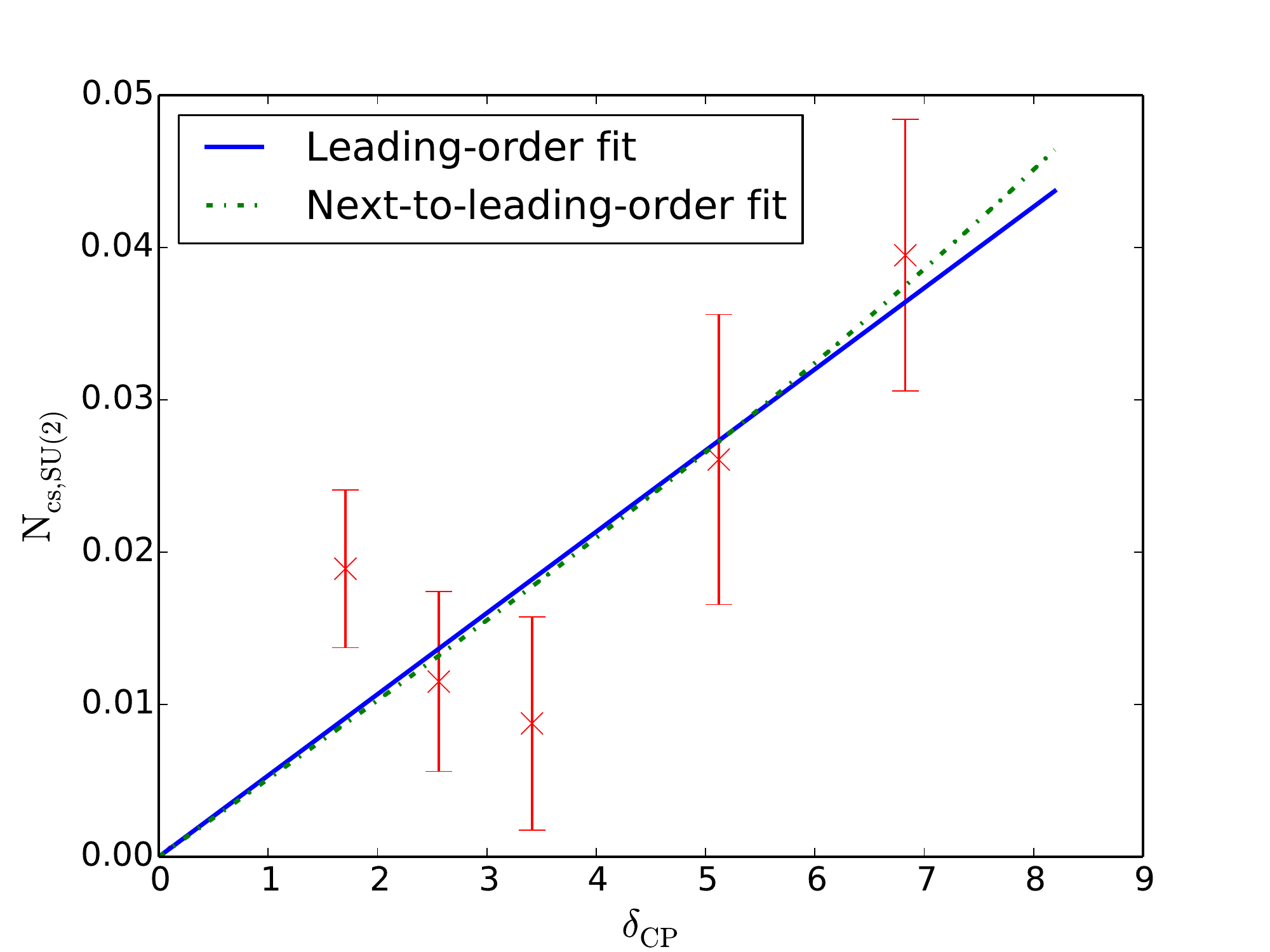} \\
\hspace{-1.5cm}
\includegraphics[width=0.6\textwidth]{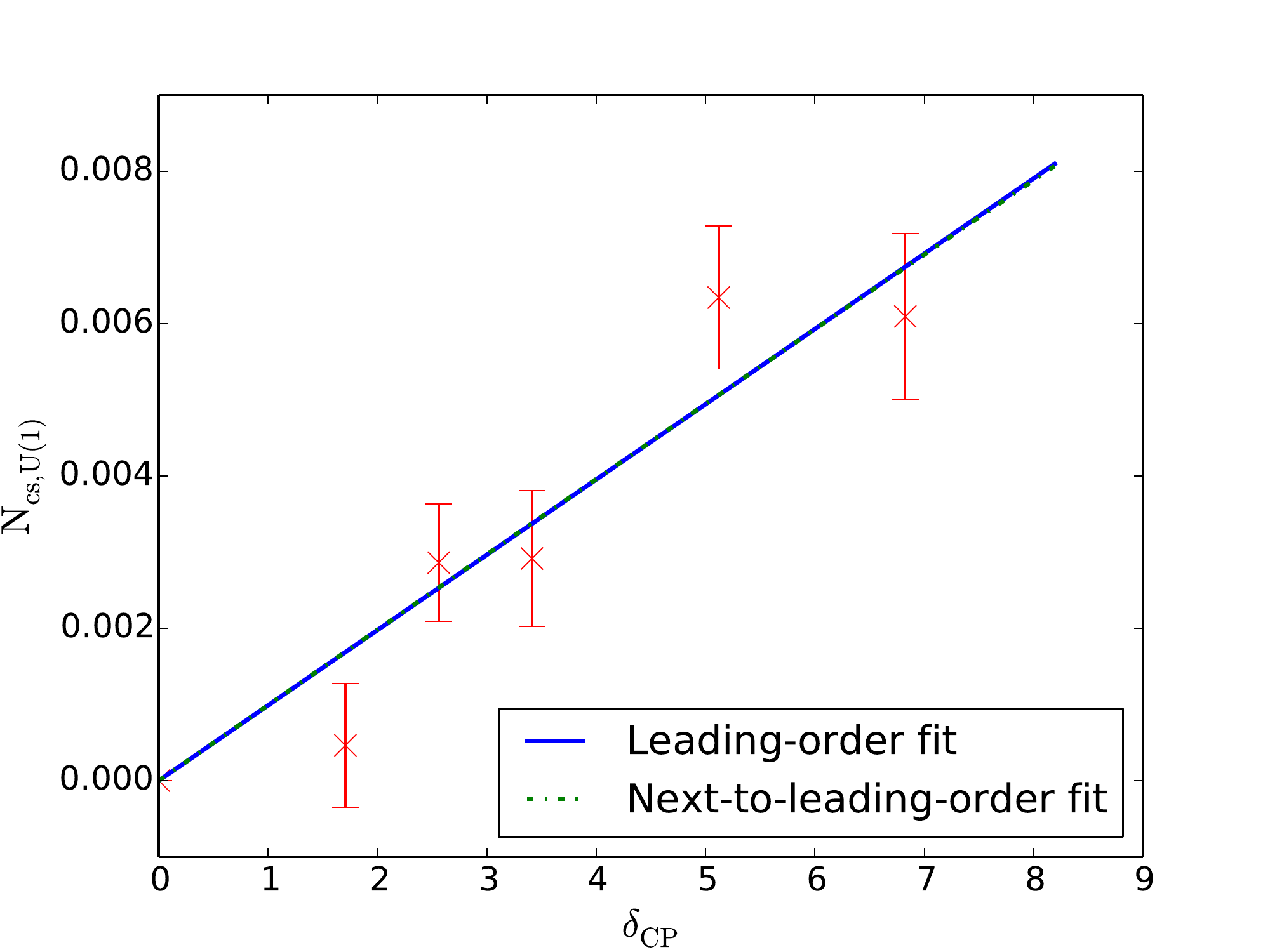} & \hspace{-1.1cm}
\includegraphics[width=0.6\textwidth]{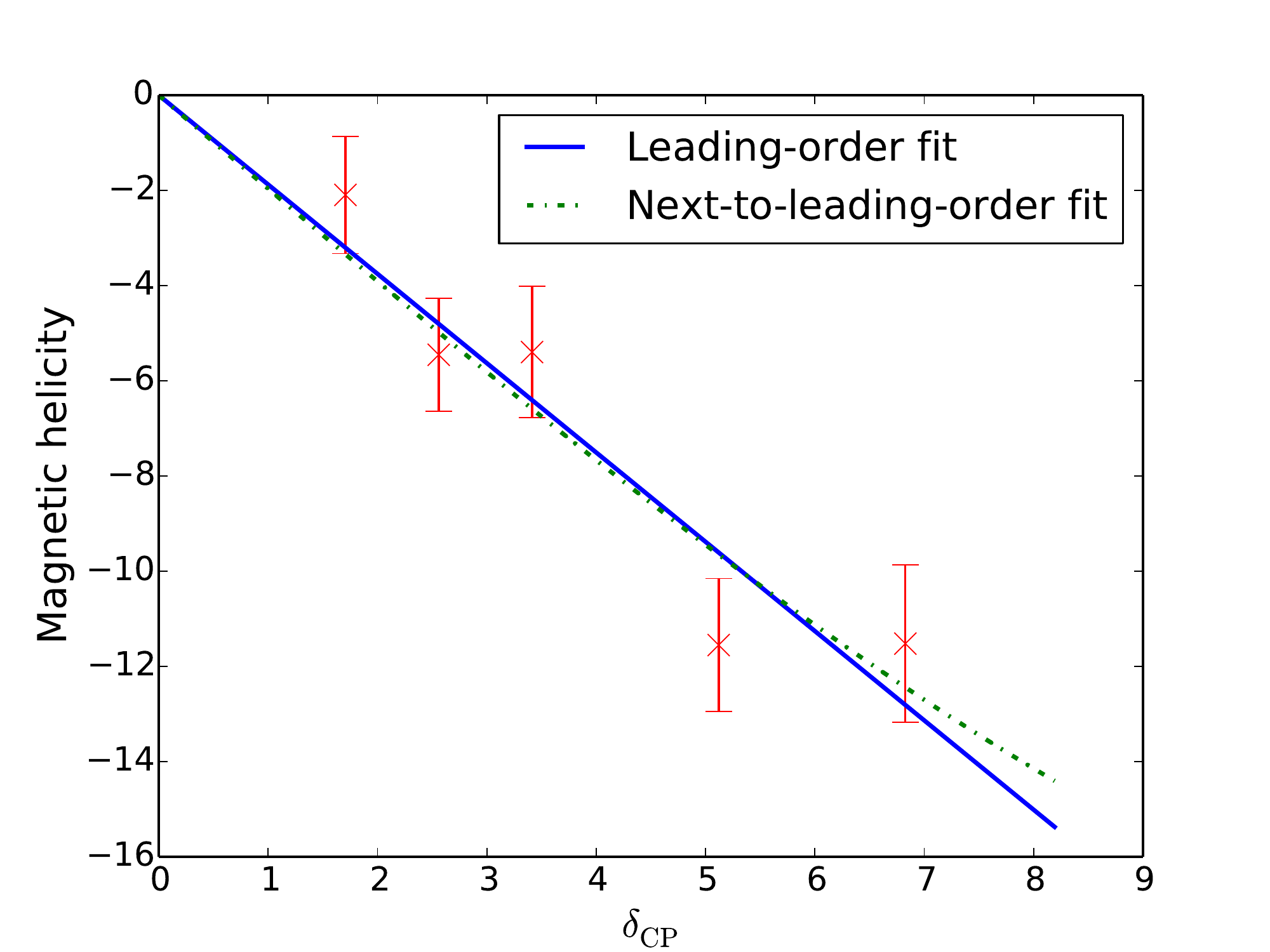}
\end{tabular}
\caption{The dependence of all the CP-odd observables on $\delta_{\rm cp}$. Clockwise from top left: $N_{\rm w}$, $N_{\rm cs,SU(2)}$, Magnetic helicity, $N_{\rm cs, U(1)}$. }
\label{fig:n2kappa}
\end{figure}

In Fig. \ref{fig:n2kappa} we first confirm the linear dependence of the asymmetry on $\delta_{\rm cp}$, using four different CP-odd observables. This is a relation established before for by-hand quenches \cite{Mou:2017atl}, but for these warmer simulations, we found it prudent to check once more. The results are taken for $n=2$, $m_H/m=4$, and are snapshots at time $m_Ht=400$. As we will discuss in detail in section \ref{sec:ncs}, this is asymptotically late for the observable $N_{\rm w}$ (top left), but not for the other CP-odd observables $N_{\rm cs, SU(2)}$ (top right), $N_{\rm cs, U(1)}$ (bottom left) and magnetic helicity\footnote{ We will not be so concerned about this observable here. Please see \cite{Mou:2017atl} for a discussion and the  precise lattice definition.}(bottom right). The dependence on the magnitude is clearly linear (blue line), and for illustration we have added the next-to-leading order fit, including a term $\propto \delta_{\rm cp}^3$ (green dashed). All other simulations in this work are performed at the largest $\delta_{\rm cp}$ included in these plots, $20/3$.

\begin{figure}
\hspace{3.cm}
\includegraphics[width=0.6\textwidth]{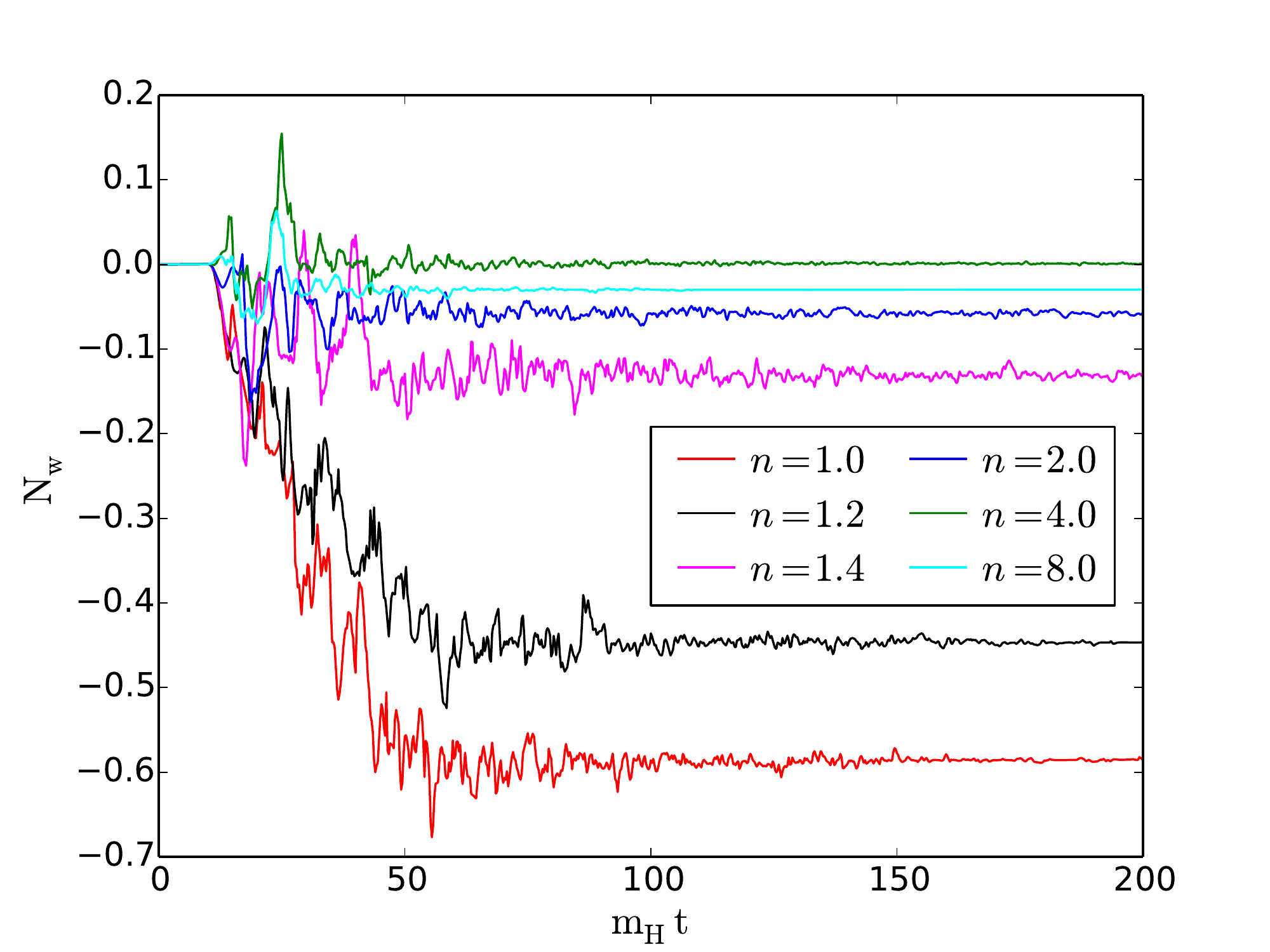}\\
\begin{tabular}{lr}
\vspace{-0.5cm}
\hspace{-1.5cm}
\includegraphics[width=0.6\textwidth]{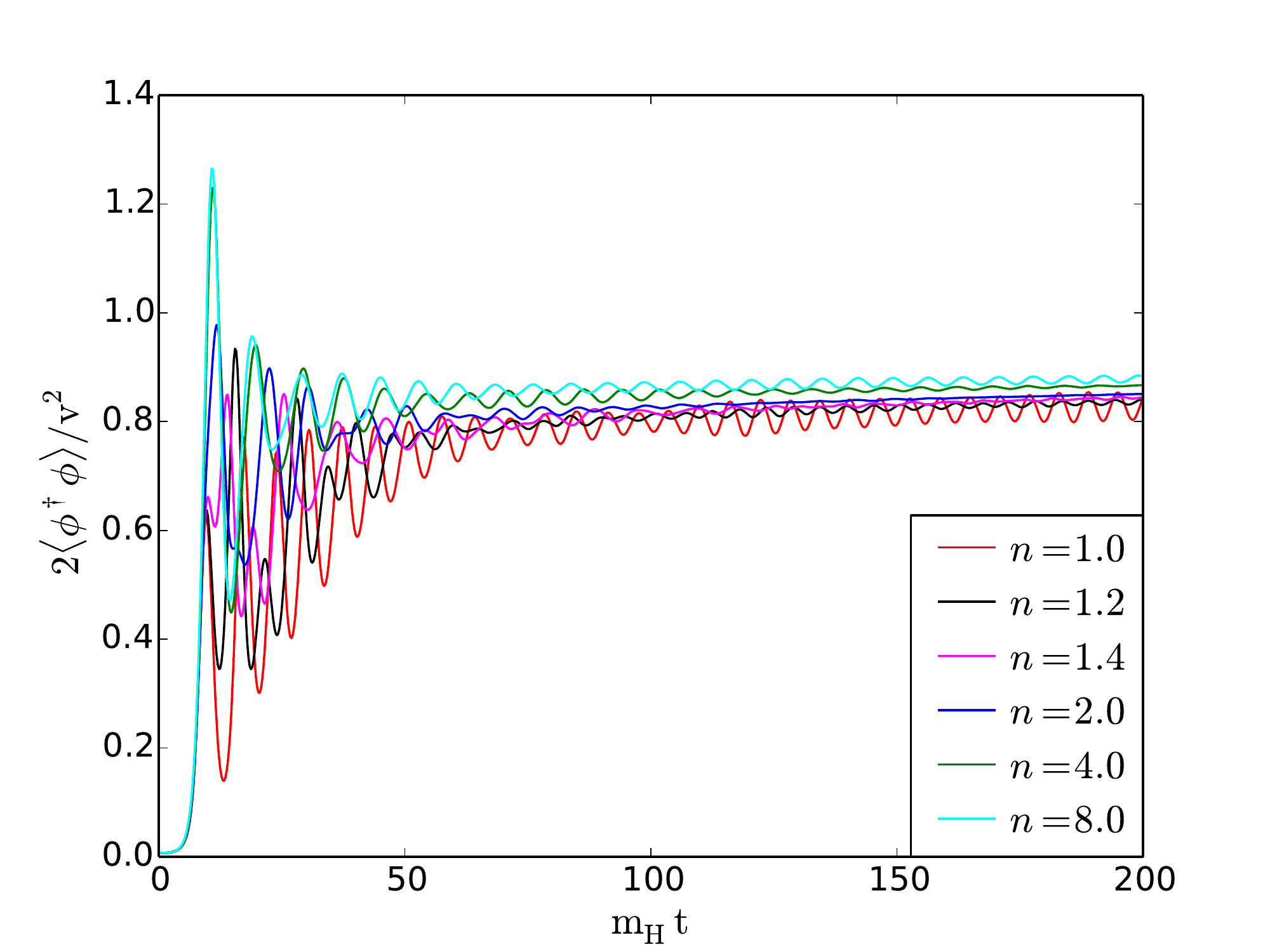} & \hspace{-1.1cm}
\includegraphics[width=0.6\textwidth]{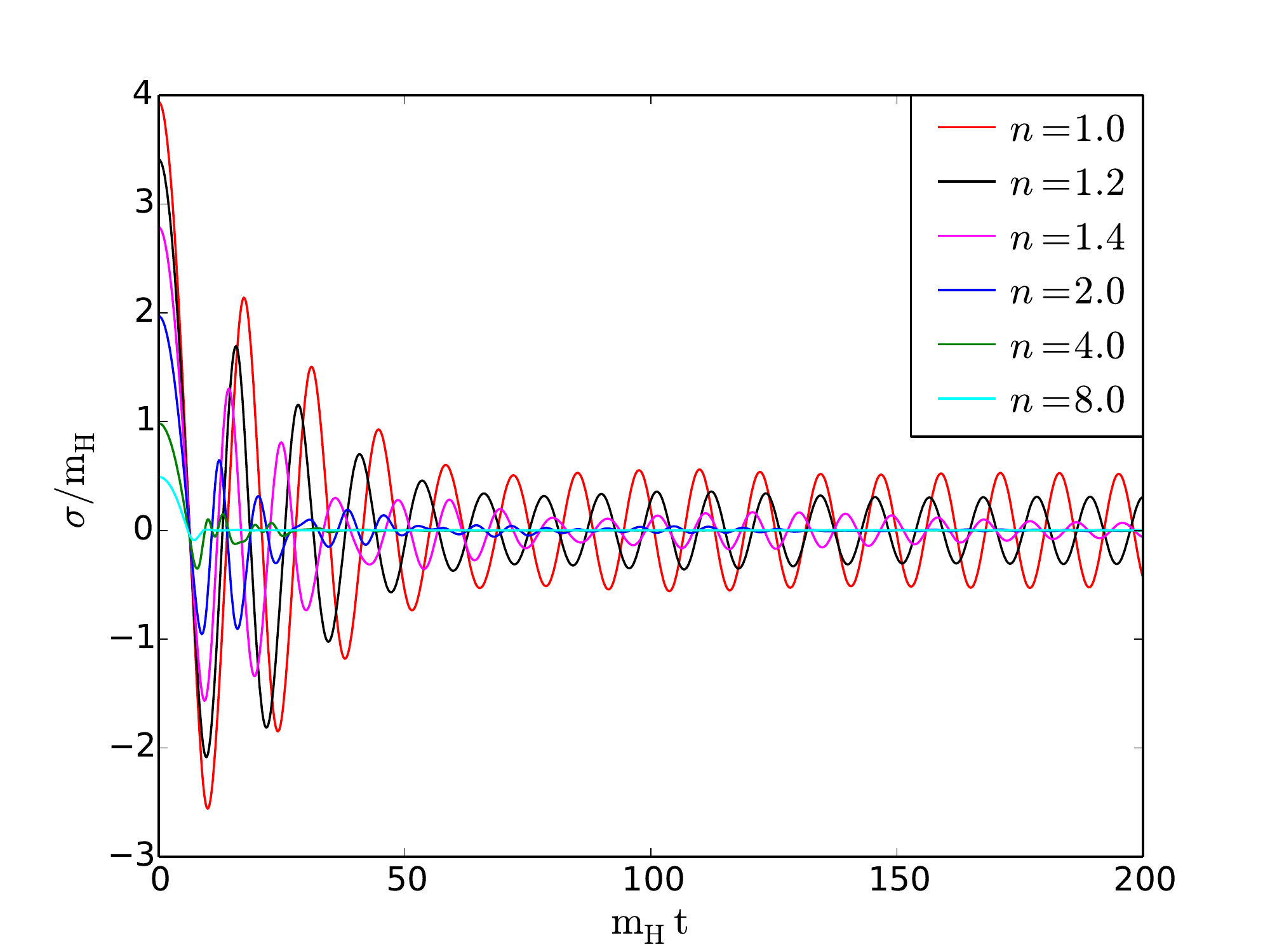}
\end{tabular}
\caption{The evolution of Higgs winding number (top), Higgs field (bottom left), and singlet field (bottom right) for different $n$. }
\label{fig:ndep1}
\end{figure}

In Fig. \ref{fig:ndep1} we show the time histories of the winding number $N_{\rm w}$ (top) and the average Higgs (bottom left) and $\sigma$ fields (bottom right) for five different $n$, at $m_H/m=4$. We see that smaller $n$ gives a larger (negative) asymmetry, and that this asymmetry is created during the first few oscillations of the Higgs field as before. 

\begin{figure}
\begin{center}
\includegraphics[width=0.7\textwidth]{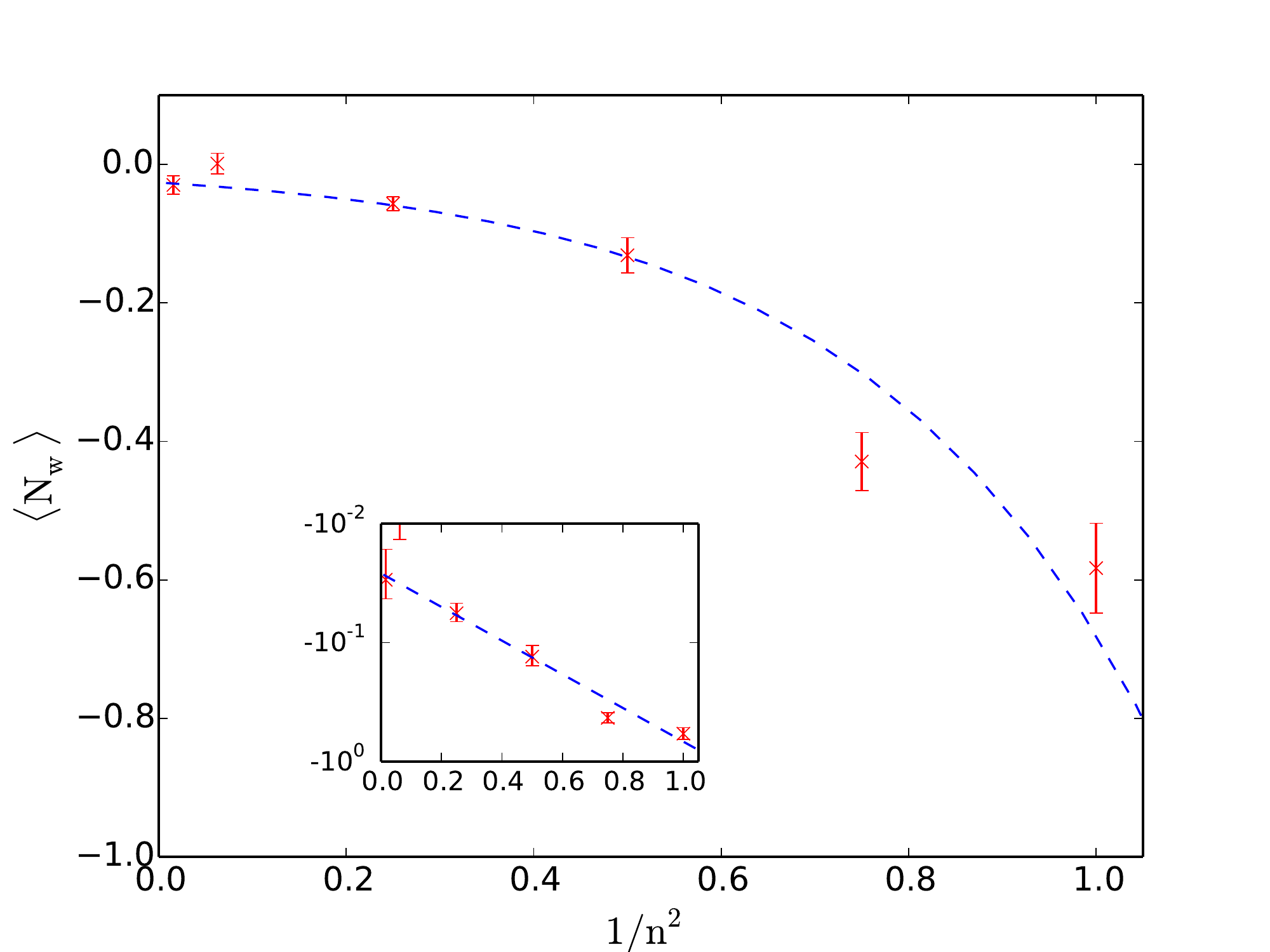}
\end{center}
\caption{The asymmetry as a function of energy in the system. Overlaid, an exponential fit (see main text). Insert: The same but on a log-scale. }
\label{fig:ndep2}
\end{figure}

Finally in Fig. \ref{fig:ndep2} we show the asymmetry as a function of $1/n^2$ (or, equivalently, $V_{\rm initial}/V_0-1$). Overlaid is an exponential fit of the form (see also insert, with a log-linear scale).
\begin{eqnarray}
\label{eq:Nw_vs_n}
\langle N_{\rm w}\rangle = (-0.026\pm 0.009) \exp\left(\frac{(3.3\pm 0.4)}{n^2}\right).
\end{eqnarray}
We see that in the limit $n\rightarrow\infty$, the asymmetry is just $N_{\rm w}=-0.026$, while for very small $n$, one may get very large asymmetries, indeed. We certainly do not expect that this exponential behaviour will continue indefinitely, but we see no reason why $1/n^2=2$ or larger would not hold, as they still represent fairly cold reheating temperatures. We are however challenged by the required numerical effort to reach such small $n$.

\section{The behaviour of $N_{\rm cs,SU(2)}$}
\label{sec:ncs}

\begin{figure}
\begin{center}
\includegraphics[width=0.7\textwidth]{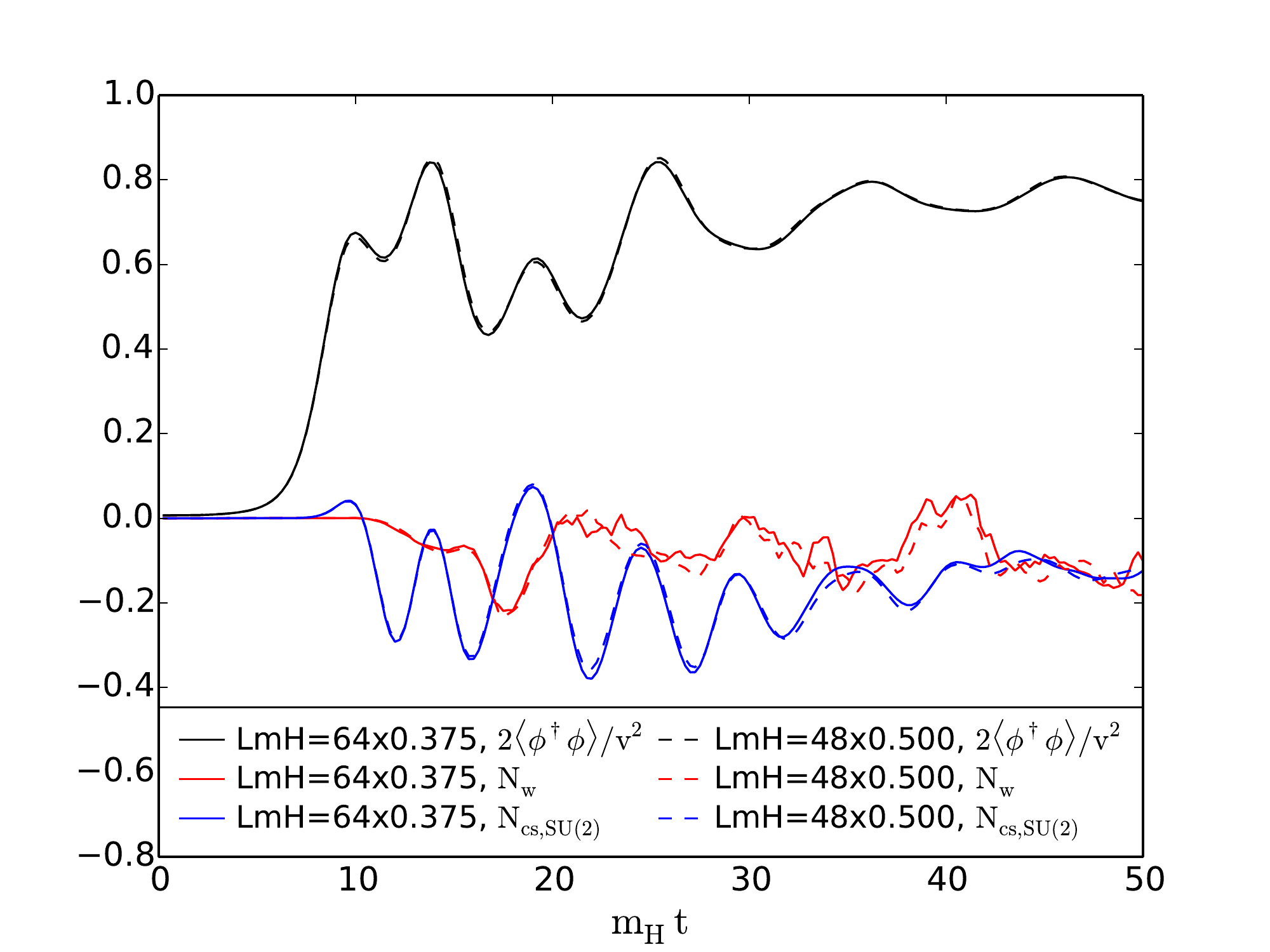}
\end{center}
\caption{The primary observables $\phi^\dagger\phi$, $N_{\rm cs,SU(2)}$, and $N_{\rm w}$ at very early times. }
\label{fig:ncs1}
\end{figure}

The chiral anomaly relates the baryon asymmetry to the SU(2) Chern-Simons number $N_{\rm cs,SU(2)}$. As described above, we have used the Higgs winding number $N_{\rm w}$ to represent the asymptotic value of the asymmetry, because dynamically it settles first, and also because it is an integer. Also recall that at low temperature, near the vacuum, the gauge field is pure gauge, and $N_{\rm cs,SU(2)}=N_{\rm w}$. We can attach a few more comments to this statement. 

In Fig. \ref{fig:ncs1} we show the early evolution of both Chern-Simons number and Higgs winding, as well as the Higgs expectation value. All observables are averaged over an initially CP-even ensemble. We see that because of the CP-violating term, $N_{\rm cs,SU(2)}$ is first biased to become positive during the transition, after which is bounces back towards a negative value. Only after this initial behaviour does the winding number change. The final asymmetry in $N_{\rm w}$ depends sensitively on the evolution of $N_{\rm cs,SU(2)}$ and on the availability of local Higgs zeros, and so on the oscillation of $\phi^\dagger\phi$ (see also the discussions in \cite{Tranberg:2006dg}).

The Higgs winding number then essentially settles, but the Chern-Simons number does not immediately drift to the same value. In fact, we see that it tends to overshoot to a larger positive value than $N_{\rm w}$. This is due to the presence of the CP-violating term, and still converging, but not yet constant, evolution of $\phi^\dagger\phi$. 

We can attempt to construct a model of this effect by postulating that the effective potential for the Chern-Simons number near a gauge-Higgs vacuum can be written in the form 
\begin{eqnarray}
\label{eq:potential}
V[N_{\rm cs,SU(2)}] = \alpha [1-\cos(2\pi N_{\rm cs,SU(2)})]-\beta \delta_{\rm cp} \dot{(\phi^\dagger\phi)}N_{\rm cs,SU(2)}.
\end{eqnarray}
The first term is the classical periodic sphaleron-like potential, with some constant $\alpha$ parametrizing the potential barrier. Along the lowest-energy path between vacua, the height of the barrier is just the sphaleron energy \cite{Klinkhamer:1984di}, $\alpha=E_{\rm sph}/2$. For a general path in configuration space, the precise value of $\alpha$ is less obvious, much less so at finite temperature or out of equilibrium. 

We get the second term in (\ref{eq:potential}) by partial integration of the CP-violating term in the action \ref{eq:S_EW}, as well as the quite strong assumption that $\phi(x)$ is homogeneous. This gives a term proportional to $N_{\rm cs,SU(2)}$ and $\dot{(\phi^\dagger\phi)}$, the size of which we will parametrize by the coefficient $\beta$ \cite{Mou:2017atl}. This means that the minimum of the potential is biased away from integer values whenever $\delta_{\rm cp}\neq 0$ and the Higgs field is not static. A fair representation of the Higgs field evolution is the form
\begin{eqnarray}
\label{eq:Higgsform}
\phi^\dagger\phi = \frac{v^2}{2}(1-e^{-\gamma t}+\epsilon \sin \tilde m t)^2,
\end{eqnarray}
for some values of $\tilde m$, $\epsilon\ll 1$ and $\gamma$. We can now proceed to find the minimum of the $N_{\rm cs,SU(2)}$ potential, by inserting (\ref{eq:Higgsform}) into (\ref{eq:potential}), to find
\begin{eqnarray}
N_{\rm cs,SU(2)}^{\rm min}= \frac{1}{2\pi}\sin^{-1}\left[
\frac{\beta \delta_{\rm cp}\tilde m v^2}{8\pi^2\alpha} e^{-2\frac{\gamma}{\tilde m}(2\pi+\tilde mt)}(1-e^{4\pi\frac{\gamma}{\tilde m}}+2e^{\frac{\gamma}{\tilde m}(2\pi+\tilde mt)}(e^{2\pi\frac{\gamma}{\tilde m}}-1)(1+\epsilon\sin(\tilde mt))
\right].\nonumber\\
\end{eqnarray}
We have averaged over one period of the Higgs field oscillation (set $t\rightarrow t+t'$, average over $t'\in[0,2\pi/\tilde m]$). Setting now $\epsilon\simeq 0$, or doing it from the beginning and not averaging, gives essentially the same result for $\tilde mt\gg 2\pi$. The expression setting $\epsilon=0$ initially leads to
\begin{eqnarray}
\label{eq:NcsAnalytic}
N_{\rm cs,SU(2)}^{\rm min}= \frac{1}{2\pi}\sin^{-1}\left[
\frac{\beta \delta_{\rm cp}\gamma v^2}{2\pi\alpha} e^{-2\gamma t}(e^{\gamma t}-1)
\right],\nonumber\\
\end{eqnarray}
The amplitude is controlled by $\delta_{\rm cp}$ and $\beta/\alpha$. The shift is substantial ($\sim0.2$), and so a linear approximation is not necessarily very accurate. This has implications for how large  $\delta_{\rm cp}$ can be allowed to be in the simulation. It should probably not be such that the intermediate-time minimum is shifted by more than $\frac{1}{2}$, since that would blur the distinction between adjacent potential minima in the original, CP-even potential.

Since the CP-even part of the potential is periodic, and $N_{\rm w}$ takes integer values for each of the ensemble configurations, we can think of the CP-violation as shifting all the minima of the $N_{\rm cs,SU(2)}$-potential away from these integer value, all in the same direction. This means that such a shift is conserved under ensemble averaging, whereas the overall asymmetry includes a cancellation between positive and negative integer flips.

\begin{figure}
\begin{center}
\includegraphics[width=0.7\textwidth]{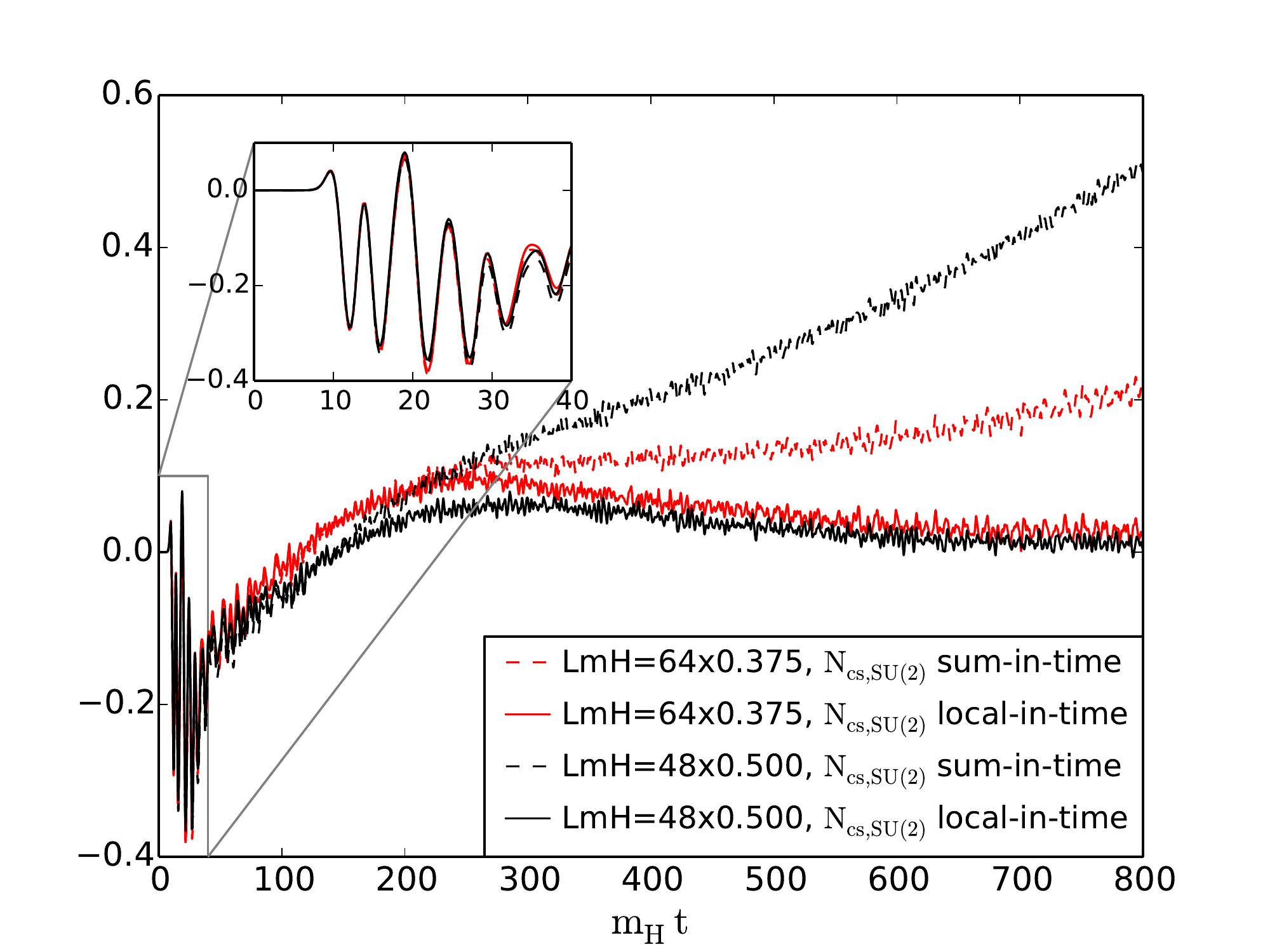}
\end{center}
\caption{The two lattice definitions of $N_{\rm cs,SU(2)}$, for two lattice spacings, with the same physical volume. The local-in-time definition performs best and is less lattice spacing dependent.}
\label{fig:ncs2}
\end{figure}

The lattice implementation used here of the observable $N_{\rm cs,SU(2)}$, is notoriously sensitive to UV fluctuations \cite{Moore:1998swa}. In equilibrium at finite temperature, it is completely essential to cool the configuration, in order to reliably measure the Chern-Simons number. For Cold Electroweak Baryogenesis, the dynamics is in the far IR modes, and the rescattering of power into the UV is quite slow \cite{Smit:2002yg,Kurkela:2012hp}. 

In Fig. \ref{fig:ncs2} we show the Chern-Simons number computed as a discretized sum in time, during the simulation
\begin{eqnarray}
N_{\rm cs,SU(2)}(t)-N_{\rm cs,SU(2)}(0) = \int dt\,d^3x \frac{1}{16\pi^2}\textrm{Tr}W_{\mu\nu}W^{\mu\nu},
\end{eqnarray}
and as a local-in-time expression
\begin{eqnarray}
N_{\rm cs,SU(2)}(t)= -\frac{g^2}{32\pi^2}\int d^3x\epsilon^{ijk}\left( W^a_i W^a_{jk}-\frac{g}{3}\epsilon^{abc}W^a_iW^b_jW^c_k \right).
\end{eqnarray}
We show this for two different lattice spacings $am_H$, $0.375$ and $0.5$, but with the same physical volume $(Lm_H)^3=24^3$. We see that computing $N_{\rm cs,SU(2)}$ without cooling is unproblematic for the first 200-250 hundred time units. For later times, a procedure based on a discretized time integral of the Chern-Simons current becomes less and less reliable, and then we must use the local-in-time approach. For even later times, 500-600, we must likely also abandon that way of calculating it, as the UV becomes populated.

\begin{figure}
\hspace{3.cm}
\includegraphics[width=0.6\textwidth]{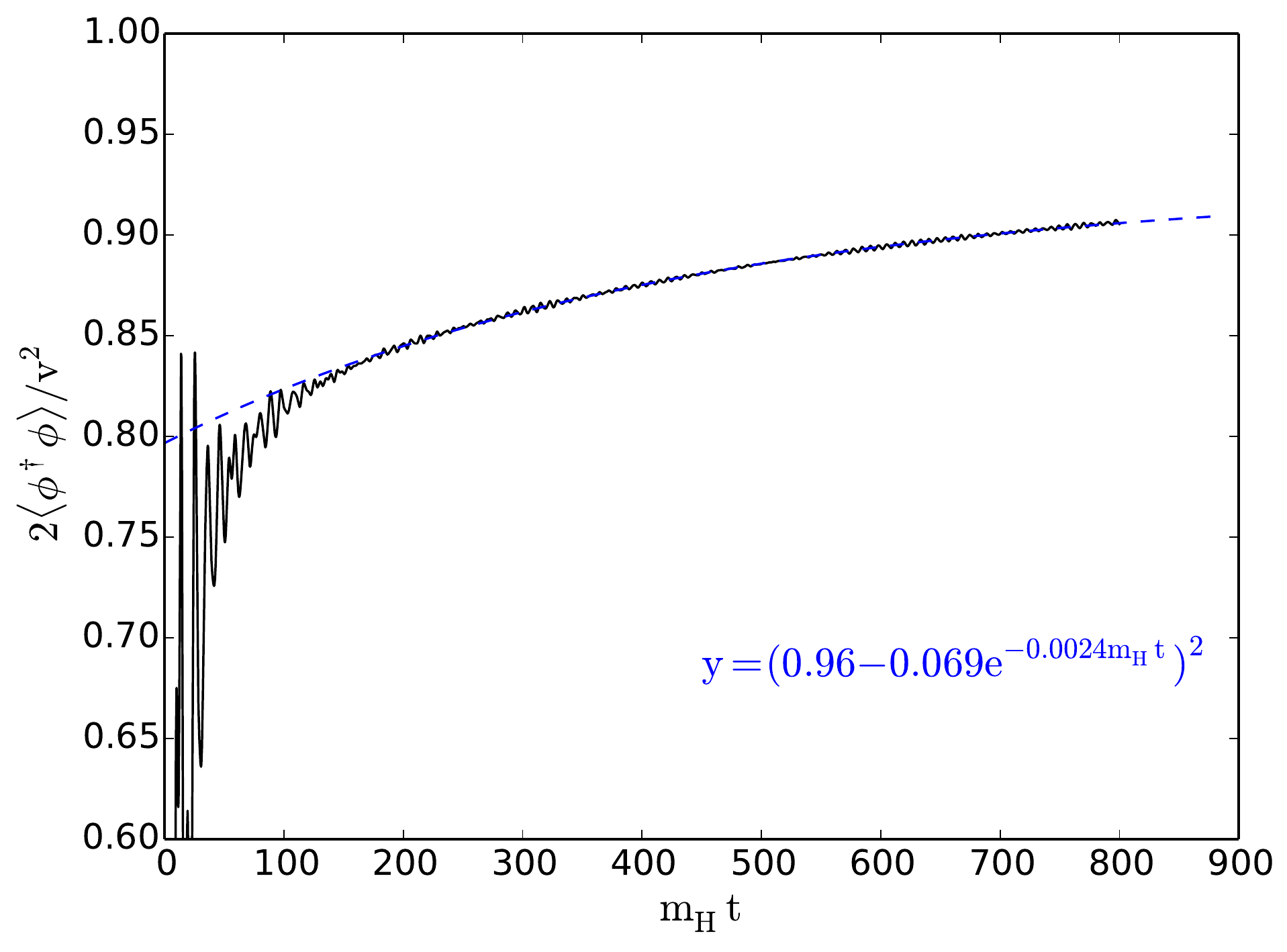}\\
\begin{tabular}{lr}
\vspace{-0.5cm}
\hspace{-1.5cm}
\includegraphics[width=0.6\textwidth]{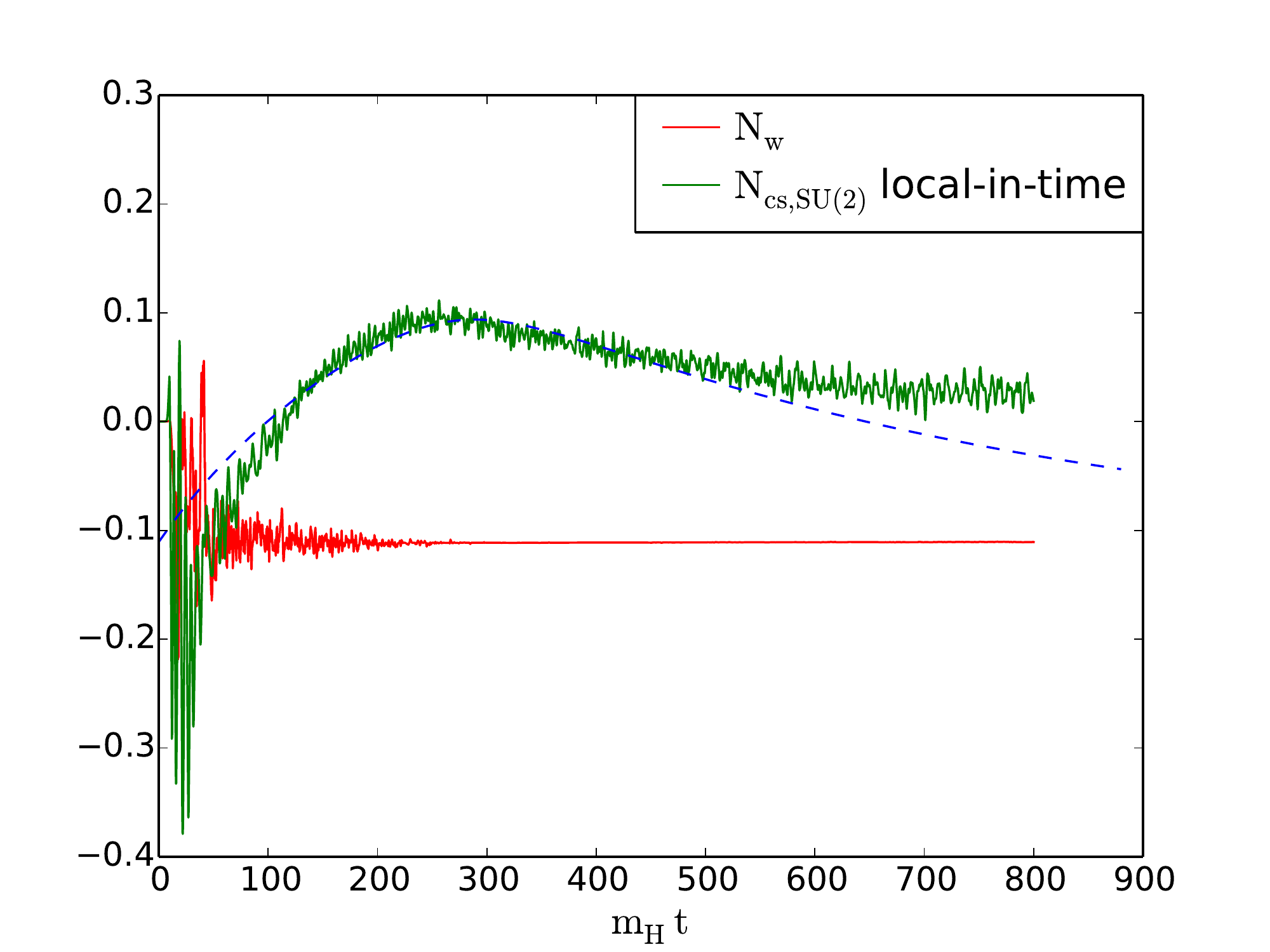} & \hspace{-1.1cm}
\includegraphics[width=0.6\textwidth]{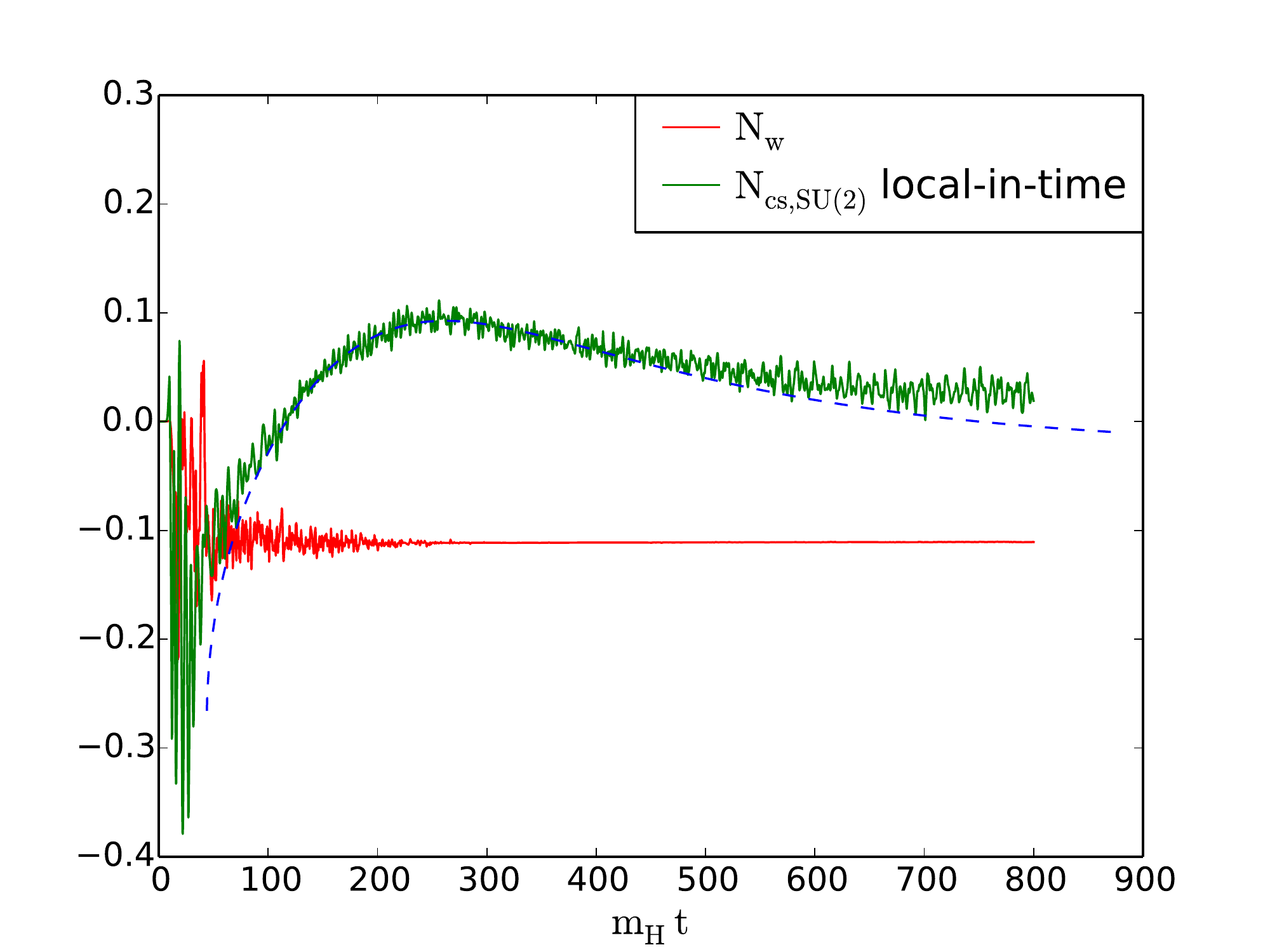}
\end{tabular}
\caption{The Higgs field appproach to the vev (top). Fits to our simple model with one free parameter (bottom left) and four free parameters (bottom right). $n=\sqrt{2}$ and $m_H/m=4$.}
\label{fig:ncs3}
\end{figure}

In Fig. \ref{fig:ncs3}, we show in the top panel the Higgs field $\phi^\dagger\phi$ as a function of time, with a fit of the form
\begin{eqnarray}
\frac{2}{v^2}\phi^\dagger\phi = \left( 0.96-0.069e^{-0.0024m_Ht} \right)^2,
\end{eqnarray}
to give us a value for the exponent $\gamma$ , which we will name $\gamma_\phi$.

We then attempt to fit $N_{\rm cs,SU(2)}$ based on the form (\ref{eq:NcsAnalytic})
\begin{eqnarray}
N_{\rm cs,SU(2)} = A+\frac{1}{2\pi}\sin^{-1}\left[Be^{-\gamma t}- Ce^{-2\gamma t}\right].
\end{eqnarray}
According to our model, we would expect $A=N_{\rm w}$, $B=C$ and $\gamma=\gamma_\phi$. It turns out to be difficult to satisfy all three constraints in a single fit, which then has only one free parameter $B=C$. Such a fit is shown in the bottom left-hand panel of Fig.~\ref{fig:ncs3}. The value of $B=C$ is $3.84$. Clearly our model is too crude to capture all the features of the dynamics. 

On the other, if we allow $A$, $B$, $C$ and $\gamma_\phi$ to be free, a much better fit is possible, shown in the bottom right-hand panel of of Fig.~\ref{fig:ncs3}. In this case we find $A=-0.024$, $B=4.18$, $C=6.53$ and $\gamma=0.0043$. Any intermediate scheme of partial fixing of parameters gives interpolating fits between the two shown. 

\begin{figure}
\begin{center}
\includegraphics[width=0.7\textwidth]{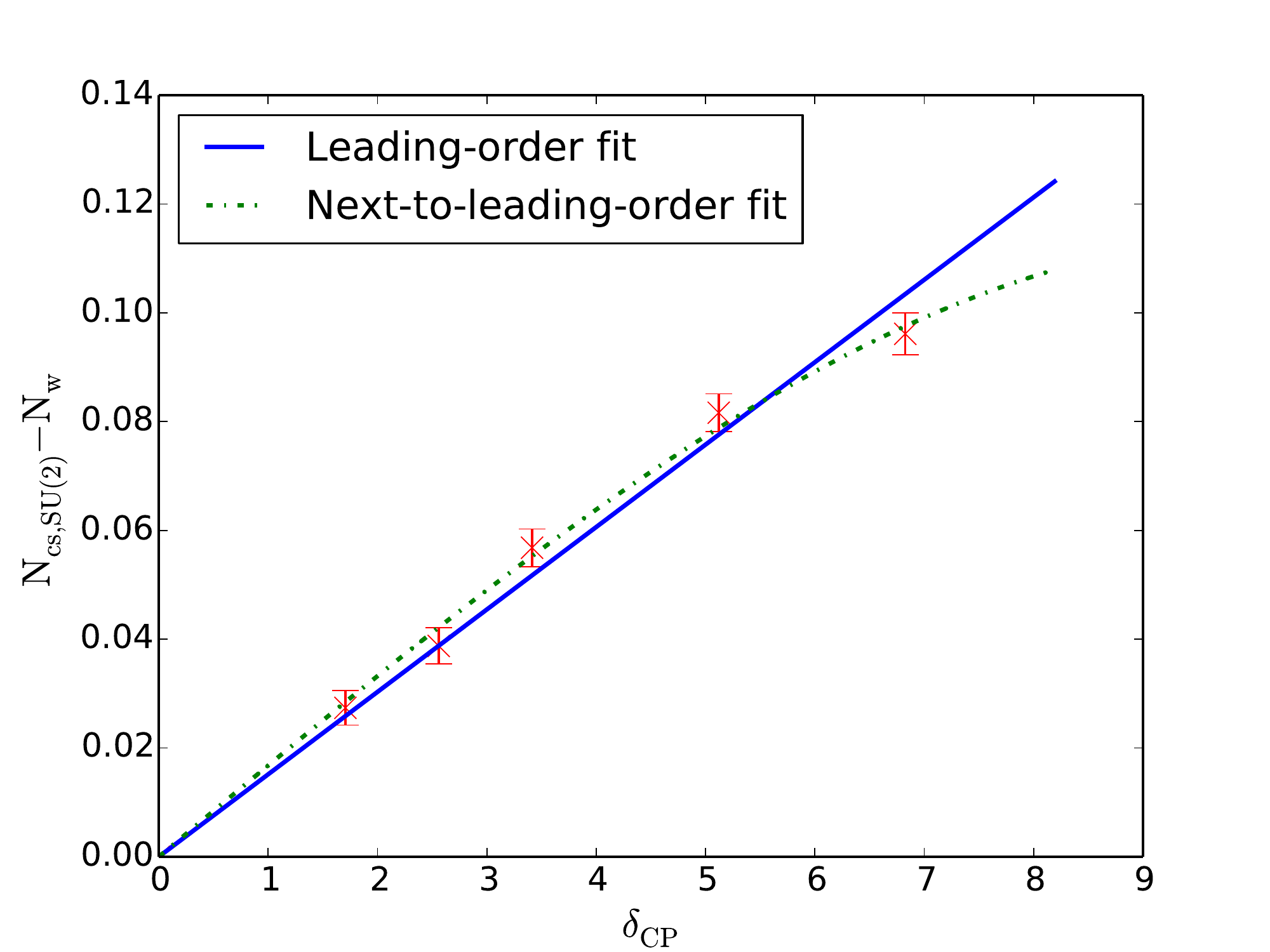}
\end{center}
\caption{The difference between $N_{\rm cs,SU(2)}$ and $N_{\rm w}$ as a function of $\delta_{\rm cp}$ at time $m_Ht=400$ with $n=2$ and $m_H/m=4$.}
\label{fig:fit_ncs_nw}
\end{figure}

One further prediction of our model, is that the shift of $N_{\rm cs,SU(2)}$ from $N_{\rm w}$ at any time later than, say $m_Ht=200$ should be approximately linear in $\delta_{\rm cp}$. In Fig.~\ref{fig:fit_ncs_nw} we show $N_{\rm cs,SU(2)}-N_{\rm w}$ at time $m_Ht=400$ as a function fo $\delta_{\rm cp}$, showing a clear linear dependence.

We conclude that we have a qualitative, and even semi-quantitative understanding of the behaviour of $N_{\rm cs,SU(2)}$ up to a time $m_Ht\simeq 500$, and that for longer times, lattice artefacts start becoming important, as power shifts into the UV. It is tempting to conclude that lattice artefacts from the UV play a dominant role for larger times. It is also possible that the coefficient $\alpha$, parametrising the depth of the sphaleron potential is time-dependent as the spectrum changes from IR-only to a more equilibrated state. We must again conclude that the time-integrated way of computing Chern-Simons number, $N_{\rm cs,t}$ is not reliable for times larger than $m_Ht=200-250$. 

We also conclude that our strategy of using $N_{\rm w}$ to represent the final asymmetry is sound, as the winding number settles completely by time $m_Ht=200$.

\section{Conclusion}
\label{sec:conc}
In this work we have examined the impact of adding a scalar singlet to the Standard Model in the context of Cold Electroweak Baryogenesis, building on earlier work where the electroweak symmetry was broken by hand over some timescale $\tau_q$ \cite{Mou:2017atl}. In the limit where the initial energy was dominated by the Higgs potential energy we were able to present a clear match between the case where the extra singlet was added, and the dynamics of the by-hand quench, finding that the quench timescale $\tau_q$ was related to the singlet mass by $\tau_q\simeq\, 1.3\,m^{-1}$, matching naive expectations.

One observation coming from the quench simulations of \cite{Tranberg:2006dg} and \cite{Mou:2017atl}  was that the final asymmetry in $N_{\rm w}$ was largest for the quench time that led to the smallest value of $\langle\phi^\dagger\phi\rangle$ during the first oscillation of the Higgs field. This was explained by noting that a small value of $\langle\phi^\dagger\phi\rangle$ at this stage allows for more Higgs-zeroes, and so increases the chances of Higgs winding events. In the simulations of this paper we have been able to extend this observation to the case where the symmetry breaking is fully dynamical, and brought about by the scalar singlet $\sigma$, finding that the asymmetry is maximised for $m_H/m\simeq30$.

From a model-building point of view, BSM scalar singlets are likely to be heavier than the Higgs field, and we therefore expect most viable realisations to generate a fast quench $m_H/m\leq 1$. In that regime we find that the asymmetry has the opposite sign compared to the slower quenches, but of the same order of magnitude. This is true for dynamical and by-hand quenches alike. 

Earlier work on the quench dynamics showed that the final Chern-Simons numbers $N_{\rm cs,SU(2)}$ depends linearly with $\delta_{\rm cp}$ \cite{Mou:2017atl}, and this also applies to the other CP-odd observables the $N_{\rm w}$, $N_{\rm cs,U(1)}$ and the magnetic helicity \cite{Mou:2017zwe}. Since these are not explicitly biased by the CP-violating term, we regard them as secondary asymmetries, sourced by their coupling to the Chern-Simons number. The simulations in this paper show that this property persists when the electroweak symmetry is broken dynamically by a singlet scalar.

Not everything is the same between the by-hand and dynamical symmetry breaking quenches. For example we find larger final $N_{\rm w}$ for the slower quenches in the simulations that use the scalar singlet, Fig. \ref{fig:asym_match}. We are also able to examine the effect on $N_{\rm w}$ of placing more of the initial energy in $\sigma$. This was done by reducing $n$ in (\ref{eq:n_energy}), with the results of Fig. \ref{fig:ndep2} showing that $N_{\rm w}$ increases exponentially, at least over the range considered, as $n$ decreases (\ref{eq:Nw_vs_n}).

We have no detailed understanding of this behaviour. In the case of equilibrium dynamics of sphaleron or sphaleron-like configurations, an exponential suppression at low temperature is natural. But here, we have an asymmetry generated by incidental flipping of the winding number, in a CP-breaking gauge field background, as a semi-coherent Higgs field oscillation produces more or less local Higgs field zeroes. The asymmetry is clearly correlated with the number of zeros, with the magnitude of CP-violation, and it seems sensible that additional energy and a faster $\sigma$ would produce a larger asymmetry. But that it would be very closely exponential is surprising. 

Future work should consider more closely the exponential dependence of $N_{\rm w}$ on $1/n^2$, as seen in Fig. \ref{fig:ndep2}. Smaller values of $n$ correspond to the scalar singlet having more energy initially, and are quite challenging numerically, but it would be interesting to see how far the exponential behaviour persists. The fact that secondary asymmetries are produced in the background of a primary asymmetry in $N_{\rm cs}$, suggests on the other hand, that a secondary asymmetry could be produced in $N_{\rm cs}$ in the case where the primary CP-violation is realised in another way (say through the $U(1)$ field). This is under investigation. 

Finally, the space of $\sigma$ initial conditions and parameters is vast, allowing for very non-linear behaviour of two-scalar oscillations. This includes cases where the $\sigma$ field oscillates with large amplitude, continually restoring and breaking the Higgs field symmetry as it passes above and below $\sigma_c$. Only as the $\sigma$ kinetic energy is transferred to the Higgs field (or itself, in the case of self-interactions) does the amplitude decrease enough that symmetry breaking completes. We have made sample runs of these, but because the phenomenology is very rich, including effects akin to parametric resonance, we postpone the detailed investigation to future work. 

\vspace{0.2cm}

\noindent
{\bf Acknowledgments:}  AT and ZGM are supported by a  UiS-ToppForsk grant from the University of Stavanger. PS acknowledges support by STFC under grant ST/L000393/1. The numerical work was performed on on the Abel Cluster, owned by the University of Oslo and the Norwegian metacenter for High Performance Computing (NOTUR), and operated by the Department for Research Computing at USIT, the University of Oslo IT-department.


\begin{thebibliography}{*}

\bibitem{Kuzmin:1985mm}
  V.~A.~Kuzmin, V.~A.~Rubakov and M.~E.~Shaposhnikov,
  Phys.\ Lett.\  {\bf 155B} (1985) 36.
  doi:10.1016/0370-2693(85)91028-7

\bibitem{Cohen:1990it} 
  A.~G.~Cohen, D.~B.~Kaplan and A.~E.~Nelson,
  Nucl.\ Phys.\ B {\bf 349}, 727 (1991).
  doi:10.1016/0550-3213(91)90395-E

\bibitem{Turok:1990in} 
  N.~Turok and J.~Zadrozny,
  Phys.\ Rev.\ Lett.\  {\bf 65}, 2331 (1990).
  doi:10.1103/PhysRevLett.65.2331

\bibitem{Cohen:1993nk}
  A.~G.~Cohen, D.~B.~Kaplan and A.~E.~Nelson,
  Ann.\ Rev.\ Nucl.\ Part.\ Sci.\  {\bf 43} (1993) 27
  doi:10.1146/annurev.ns.43.120193.000331
  [hep-ph/9302210].

\bibitem{tHooft:1976rip} 
  G.~'t Hooft,
  Phys.\ Rev.\ Lett.\  {\bf 37}, 8 (1976).
  doi:10.1103/PhysRevLett.37.8

\bibitem{tHooft:1976snw} 
  G.~'t Hooft,
  Phys.\ Rev.\ D {\bf 14}, 3432 (1976)
  Erratum: [Phys.\ Rev.\ D {\bf 18}, 2199 (1978)].
  doi:10.1103/PhysRevD.18.2199.3, 10.1103/PhysRevD.14.3432

\bibitem{Choi:1993cv} 
  J.~Choi and R.~R.~Volkas,
  Phys.\ Lett.\ B {\bf 317}, 385 (1993)
  doi:10.1016/0370-2693(93)91013-D
  [hep-ph/9308234].

\bibitem{Huber:2006wf} 
  S.~J.~Huber, T.~Konstandin, T.~Prokopec and M.~G.~Schmidt,
  Nucl.\ Phys.\ B {\bf 757}, 172 (2006)
  doi:10.1016/j.nuclphysb.2006.09.003
  [hep-ph/0606298].

\bibitem{Cheung:2012pg} 
  K.~Cheung, T.~J.~Hou, J.~S.~Lee and E.~Senaha,
  Phys.\ Lett.\ B {\bf 710}, 188 (2012)
  doi:10.1016/j.physletb.2012.02.070
  [arXiv:1201.3781 [hep-ph]].

\bibitem{Espinosa:2011ax} 
  J.~R.~Espinosa, T.~Konstandin and F.~Riva,
  Nucl.\ Phys.\ B {\bf 854}, 592 (2012)
  doi:10.1016/j.nuclphysb.2011.09.010
  [arXiv:1107.5441 [hep-ph]].

\bibitem{Damgaard:2015con} 
  P.~H.~Damgaard, A.~Haarr, D.~O'Connell and A.~Tranberg,
  JHEP {\bf 1602}, 107 (2016)
  doi:10.1007/JHEP02(2016)107
  [arXiv:1512.01963 [hep-ph]].

\bibitem{Alanne:2016wtx} 
  T.~Alanne, K.~Kainulainen, K.~Tuominen and V.~Vaskonen,
  JCAP {\bf 1608}, no. 08, 057 (2016)
  doi:10.1088/1475-7516/2016/08/057
  [arXiv:1607.03303 [hep-ph]].

\bibitem{Morrissey:2012db} 
  D.~E.~Morrissey and M.~J.~Ramsey-Musolf,
  New J.\ Phys.\  {\bf 14}, 125003 (2012)
  doi:10.1088/1367-2630/14/12/125003
  [arXiv:1206.2942 [hep-ph]].

\bibitem{Krauss:1999ng}
  L.~M.~Krauss and M.~Trodden,
  Phys.\ Rev.\ Lett.\  {\bf 83} (1999) 1502
  doi:10.1103/PhysRevLett.83.1502
  [hep-ph/9902420].

\bibitem{GarciaBellido:1999sv}
  J.~Garcia-Bellido, D.~Y.~Grigoriev, A.~Kusenko and M.~E.~Shaposhnikov,
  Phys.\ Rev.\ D {\bf 60} (1999) 123504
  doi:10.1103/PhysRevD.60.123504
  [hep-ph/9902449].

\bibitem{Rajantie:2000nj}
  A.~Rajantie, P.~M.~Saffin and E.~J.~Copeland,
  Phys.\ Rev.\ D {\bf 63} (2001) 123512
  doi:10.1103/PhysRevD.63.123512
  [hep-ph/0012097].

\bibitem{Copeland:2001qw}
  E.~J.~Copeland, D.~Lyth, A.~Rajantie and M.~Trodden,
  Phys.\ Rev.\ D {\bf 64} (2001) 043506
  doi:10.1103/PhysRevD.64.043506
  [hep-ph/0103231].

\bibitem{Smit:2002yg}
  J.~Smit and A.~Tranberg,
  JHEP {\bf 0212} (2002) 020
  doi:10.1088/1126-6708/2002/12/020
  [hep-ph/0211243].

\bibitem{Tranberg:2006ip}
  A.~Tranberg and J.~Smit,
  JHEP {\bf 0608} (2006) 012
  doi:10.1088/1126-6708/2006/08/012
  [hep-ph/0604263].

\bibitem{Konstandin:2011ds}
  T.~Konstandin and G.~Servant,
  JCAP {\bf 1107} (2011) 024
  doi:10.1088/1475-7516/2011/07/024
  [arXiv:1104.4793 [hep-ph]].

\bibitem{Enqvist:2010fd}
  K.~Enqvist, P.~Stephens, O.~Taanila and A.~Tranberg,
  JCAP {\bf 1009} (2010) 019
  doi:10.1088/1475-7516/2010/09/019
  [arXiv:1005.0752 [astro-ph.CO]].

\bibitem{vanTent:2004rc}
  B.~J.~W.~van Tent, J.~Smit and A.~Tranberg,
  JCAP {\bf 0407} (2004) 003
  doi:10.1088/1475-7516/2004/07/003
  [hep-ph/0404128].

\bibitem{Mou:2015aia}
  Z.~G.~Mou, P.~M.~Saffin and A.~Tranberg,
  JHEP {\bf 1506} (2015) 163
  doi:10.1007/JHEP06(2015)163
  [arXiv:1505.02692 [hep-ph]].

\bibitem{Tranberg:2012qu}
  A.~Tranberg and B.~Wu,
  JHEP {\bf 1301} (2013) 046
  doi:10.1007/JHEP01(2013)046
  [arXiv:1210.1779 [hep-ph]].

\bibitem{Tranberg:2012jp}
  A.~Tranberg and B.~Wu,
  JHEP {\bf 1207} (2012) 087
  doi:10.1007/JHEP07(2012)087
  [arXiv:1203.5012 [hep-ph]].

\bibitem{Tranberg:2003gi}
  A.~Tranberg and J.~Smit,
  JHEP {\bf 0311} (2003) 016
  doi:10.1088/1126-6708/2003/11/016
  [hep-ph/0310342].

\bibitem{Tranberg:2006dg}
  A.~Tranberg, J.~Smit and M.~Hindmarsh,
  JHEP {\bf 0701} (2007) 034
  doi:10.1088/1126-6708/2007/01/034
  [hep-ph/0610096].

\bibitem{vanderMeulen:2005sp}
  M.~van der Meulen, D.~Sexty, J.~Smit and A.~Tranberg,
  JHEP {\bf 0602} (2006) 029
  doi:10.1088/1126-6708/2006/02/029
  [hep-ph/0511080].


\bibitem{GarciaBellido:2003wd}
  J.~Garcia-Bellido, M.~Garcia-Perez and A.~Gonzalez-Arroyo,
  Phys.\ Rev.\ D {\bf 69} (2004) 023504
  doi:10.1103/PhysRevD.69.023504
  [hep-ph/0304285].

\bibitem{Mou:2017atl} 
  Z.~G.~Mou, P.~M.~Saffin and A.~Tranberg,
  JHEP {\bf 1707}, 010 (2017)
  doi:10.1007/JHEP07(2017)010
  [arXiv:1703.01781 [hep-ph]].

\bibitem{Mou:2017zwe} 
  Z.~G.~Mou, P.~M.~Saffin and A.~Tranberg,
  JHEP {\bf 1706}, 075 (2017)
  doi:10.1007/JHEP06(2017)075
  [arXiv:1704.08888 [hep-ph]].

\bibitem{Brauner:2011vb}
  T.~Brauner, O.~Taanila, A.~Tranberg and A.~Vuorinen,
  Phys.\ Rev.\ Lett.\  {\bf 108} (2012) 041601
  doi:10.1103/PhysRevLett.108.041601
  [arXiv:1110.6818 [hep-ph]].

\bibitem{Brauner:2012gu} 
  T.~Brauner, O.~Taanila, A.~Tranberg and A.~Vuorinen,
  JHEP {\bf 1211}, 076 (2012)
  doi:10.1007/JHEP11(2012)076
  [arXiv:1208.5609 [hep-ph]].

\bibitem{GarciaBellido:2002aj}
  J.~Garcia-Bellido, M.~Garcia Perez and A.~Gonzalez-Arroyo,
  Phys.\ Rev.\ D {\bf 67} (2003) 103501
  doi:10.1103/PhysRevD.67.103501
  [hep-ph/0208228].
    
 \bibitem{DiazGil:2007dy}
  A.~Diaz-Gil, J.~Garcia-Bellido, M.~Garcia Perez and A.~Gonzalez-Arroyo,
  Phys.\ Rev.\ Lett.\  {\bf 100} (2008) 241301
  doi:10.1103/PhysRevLett.100.241301
  [arXiv:0712.4263 [hep-ph]].

\bibitem{DiazGil:2008tf}
  A.~Diaz-Gil, J.~Garcia-Bellido, M.~Garcia Perez and A.~Gonzalez-Arroyo,
  JHEP {\bf 0807} (2008) 043
  doi:10.1088/1126-6708/2008/07/043
  [arXiv:0805.4159 [hep-ph]].


\bibitem{Damgaard:2013kva}
  P.~H.~Damgaard, D.~O'Connell, T.~C.~Petersen and A.~Tranberg,
  Phys.\ Rev.\ Lett.\  {\bf 111} (2013) no.22,  221804
  doi:10.1103/PhysRevLett.111.221804
  [arXiv:1305.4362 [hep-ph]].

\bibitem{Starobinsky:1986fx} 
  A.~A.~Starobinsky,
  Lect.\ Notes Phys.\  {\bf 246}, 107 (1986).
  doi:10.1007/3-540-16452-9\_6

\bibitem{Nakao:1988yi} 
  K.~i.~Nakao, Y.~Nambu and M.~Sasaki,
  Prog.\ Theor.\ Phys.\  {\bf 80}, 1041 (1988).
  doi:10.1143/PTP.80.1041

\bibitem{Stewart:1991dy} 
  J.~M.~Stewart,
  Class.\ Quant.\ Grav.\  {\bf 8}, 909 (1991).
  doi:10.1088/0264-9381/8/5/015

\bibitem{Starobinsky:1994bd} 
  A.~A.~Starobinsky and J.~Yokoyama,
  Phys.\ Rev.\ D {\bf 50}, 6357 (1994)
  doi:10.1103/PhysRevD.50.6357
  [astro-ph/9407016].

\bibitem{Enqvist:2012xn} 
  K.~Enqvist, R.~N.~Lerner, O.~Taanila and A.~Tranberg,
  JCAP {\bf 1210}, 052 (2012)
  doi:10.1088/1475-7516/2012/10/052
  [arXiv:1205.5446 [astro-ph.CO]].

\bibitem{Kajantie:1996mn} 
  K.~Kajantie, M.~Laine, K.~Rummukainen and M.~E.~Shaposhnikov,
  Phys.\ Rev.\ Lett.\  {\bf 77}, 2887 (1996)
  doi:10.1103/PhysRevLett.77.2887
  [hep-ph/9605288].

\bibitem{Karsch:1996yh} 
  F.~Karsch, T.~Neuhaus, A.~Patkos and J.~Rank,
  Nucl.\ Phys.\ Proc.\ Suppl.\  {\bf 53}, 623 (1997)
  doi:10.1016/S0920-5632(96)00736-0
  [hep-lat/9608087].

\bibitem{Aoki:1996cu} 
  Y.~Aoki,
  Phys.\ Rev.\ D {\bf 56}, 3860 (1997)
  doi:10.1103/PhysRevD.56.3860
  [hep-lat/9612023].

\bibitem{Gurtler:1997hr} 
  M.~Gurtler, E.~M.~Ilgenfritz and A.~Schiller,
  Phys.\ Rev.\ D {\bf 56}, 3888 (1997)
  doi:10.1103/PhysRevD.56.3888
  [hep-lat/9704013].

\bibitem{Laine:1998jb} 
  M.~Laine and K.~Rummukainen,
  Nucl.\ Phys.\ Proc.\ Suppl.\  {\bf 73}, 180 (1999)
  doi:10.1016/S0920-5632(99)85017-8
  [hep-lat/9809045].

\bibitem{DOnofrio:2014rug}
  M.~D'Onofrio, K.~Rummukainen and A.~Tranberg,
  Phys.\ Rev.\ Lett.\  {\bf 113} (2014) no.14,  141602
  doi:10.1103/PhysRevLett.113.141602
  [arXiv:1404.3565 [hep-ph]].

\bibitem{Laine:2015kra} 
  M.~Laine and M.~Meyer,
  JCAP {\bf 1507}, no. 07, 035 (2015)
  doi:10.1088/1475-7516/2015/07/035
  [arXiv:1503.04935 [hep-ph]].

\bibitem{Saffin:2011kn} 
  P.~M.~Saffin and A.~Tranberg,
  JHEP {\bf 1202}, 102 (2012)
  doi:10.1007/JHEP02(2012)102
  [arXiv:1111.7136 [hep-ph]].

\bibitem{GarciaRecio:2009zp} 
  C.~Garcia-Recio and L.~L.~Salcedo,
  JHEP {\bf 0907}, 015 (2009)
  doi:10.1088/1126-6708/2009/07/015
  [arXiv:0903.5494 [hep-ph]].

\bibitem{Hernandez:2008db} 
  A.~Hernandez, T.~Konstandin and M.~G.~Schmidt,
  Nucl.\ Phys.\ B {\bf 812}, 290 (2009)
  doi:10.1016/j.nuclphysb.2008.12.021
  [arXiv:0810.4092 [hep-ph]].

\bibitem{Tranberg:2010af} 
  A.~Tranberg,
  Phys.\ Rev.\ D {\bf 84}, 083516 (2011)
  doi:10.1103/PhysRevD.84.083516
  [arXiv:1009.2358 [hep-ph]].

\bibitem{Klinkhamer:1984di} 
  F.~R.~Klinkhamer and N.~S.~Manton,
  Phys.\ Rev.\ D {\bf 30}, 2212 (1984).
  doi:10.1103/PhysRevD.30.2212

\bibitem{Moore:1998swa}
  G.~D.~Moore,
  Phys.\ Rev.\ D {\bf 59} (1999) 014503
  doi:10.1103/PhysRevD.59.014503
  [hep-ph/9805264].

\bibitem{Kurkela:2012hp} 
  A.~Kurkela and G.~D.~Moore,
  Phys.\ Rev.\ D {\bf 86}, 056008 (2012)
  doi:10.1103/PhysRevD.86.056008
  [arXiv:1207.1663 [hep-ph]].

\end{thebibliography}
\end{document}